\documentclass[twocolumn,showpacs,aps,pre]{revtex4-1}
\usepackage{amssymb,amsmath}
\usepackage[dvips]{graphicx}

\begin{document}

\title{Contact process with temporal disorder}

\author{Hatem Barghathi}
\affiliation{Department of Physics, Missouri University of Science and Technology,
Rolla, MO 65409, USA}

\author{Jos\'e A. Hoyos}
\affiliation{Instituto de F\'{i}sica de S\~ao Carlos, Universidade de S\~ao Paulo,
C.P. 369, S\~ao Carlos, S\~ao Paulo 13560-970, Brazil}

\author{Thomas Vojta}
\affiliation{Department of Physics, Missouri University of Science and Technology,
Rolla, MO 65409, USA}

\begin{abstract}
We investigate the influence of time-varying environmental noise, i.e., temporal disorder,
on the nonequilibrium phase transition of the contact process. Combining a real-time
renormalization group, scaling theory, and large scale Monte-Carlo simulations in one and two dimensions,
we show that the temporal disorder gives rise to an exotic critical point. At criticality, the effective
noise amplitude diverges with increasing time scale, and the probability distribution of the
density becomes infinitely broad, even on a logarithmic scale. Moreover, the average density
and survival probability decay only logarithmically with time. This infinite-noise critical
behavior can be understood as the temporal counterpart of infinite-randomness critical behavior
in spatially disordered systems, but with exchanged roles of space and time. We also
analyze the generality of our results, and we discuss potential experiments.
\end{abstract}

\date{\today}

\maketitle

\section{Introduction}
\label{sec:Intro}

Directed percolation (DP) is the prototypical universality class of nonequilibrium phase transitions
between active, fluctuating states and fluctuationless absorbing states.
According to a conjecture by Janssen and Grassberger \cite{Janssen81,Grassberger82},
all absorbing state transitions with a scalar order parameter, short-range interactions,
and no extra symmetries or conservation laws belong to this class. DP
critical behavior has been predicted to occur, for example, in the contact process \cite{HarrisTE74},
catalytic chemical reactions \cite{ZiffGulariBarshad86}, interface growth \cite{TangLeschhorn92}
and dynamics \cite{BarabasiGrinsteinMunoz96}, as well as in turbulence \cite{Pomeau86}
(see also Refs.\ \cite{MarroDickman99,Hinrichsen00,Odor04,Luebeck04,TauberHowardVollmayrLee05,HenkelHinrichsenLuebeck_book08}
for reviews).

Despite its ubiquity in theory and computer simulations, experimental observations of DP critical behavior
were lacking for a long time \cite{Hinrichsen00b}. A full verification of this universality class
was achieved in the transition between two turbulent states in a liquid crystal
\cite{TKCS07}. Other examples of  experimental systems undergoing absorbing state transitions
include periodically driven suspensions \cite{CCGP08,FFGP11}, superconducting vortices \cite{OkumaTsugawaMotohashi11},
and bacteria colony biofilms \cite{KorolevNelson11,KXNF11}.

One of the reasons for the rarity of DP behavior in experiments is likely
the presence of disorder in most realistic systems. In fact, the DP critical point
is unstable against spatial disorder because its correlation length exponent $\nu_\perp$ violates
the Harris criterion \cite{Harris74} $d\nu_\perp>2$  in all physical dimensions. Along the same lines,
the DP critical point is unstable against temporal disorder because its correlation
time exponent $\nu_\parallel=z\nu_\perp$ violates Kinzel's generalization \cite{Kinzel85}
$\nu_\parallel > 2$ of the Harris criterion (see Ref.\ \cite{VojtaDickman16} for the stability
with respect to general spatio-temporal disorder).

The effects of spatial disorder on the DP universality class have been studied in detail using both analytical and numerical
approaches. Hooyberghs et al.\ \cite{HooyberghsIgloiVanderzande03,*HooyberghsIgloiVanderzande04}
implemented a strong-disorder renormalization group (RG) \cite{MaDasguptaHu79,IgloiMonthus05} for the
disordered contact process and predicted that the transition
is controlled by an exotic infinite-randomness critical point (at least for sufficiently strong disorder
\footnote{A self-consistent extension of the method to the weak-disorder regime is presented in Ref.\ \cite{Hoyos08}.}),
 accompanied by strong power-law
Griffiths singularities \cite{Griffiths69,Noest86,*Noest88}.
The infinite-randomness critical point was confirmed by extensive Monte-Carlo simulations in one, two and three space dimensions
\cite{VojtaDickison05,OliveiraFerreira08,VojtaFarquharMast09,Vojta12}. Analogous behavior was found
in diluted systems close to the percolation threshold \cite{VojtaLee06,*LeeVojta09} and in quasiperiodic systems
\cite{BarghathiNozadzeVojta14} (for a review, see Ref.\ \cite{Vojta06}).

The fate of the DP transition under the influence of temporal disorder, i.e., environmental noise,
has received less attention so far. Jensen applied Monte-Carlo simulations \cite{Jensen96}
and series expansions \cite{Jensen96,Jensen05} to directed bond percolation with temporal disorder.
He reported power-law scaling, but with nonuniversal exponents that change continuously with
the disorder strength. (Note that Jensen's values for the correlation time exponent $\nu_\parallel$
violate Kinzel's bound $\nu_\parallel > 2$ for weaker disorder.)
Vazquez et al.\ \cite{VBLM11} revisited this problem focusing on the effects of rare
strong fluctuations of the temporal disorder. They identified a temporal analog of the Griffiths
phase in spatially disordered systems that features an unusual power-law relation between lifetime
and system size. Recently, Vojta and Hoyos developed a real-time strong-noise RG
\cite{VojtaHoyos15} for the temporally disordered contact process. This method predicts
an exotic infinite-noise critical point at which the effective
disorder strength diverges with increasing time scale. The probability distribution of the
density becomes infinitely broad, even on a logarithmic scale,  and the average density
and survival probability at criticality decay only logarithmically with time.

In the present paper, we employ large-scale Monte-Carlo simulations to test the predictions of this RG theory.
We study the contact process with temporal disorder in one and two space dimensions
performing both spreading and density decay simulations; and we analyze the numerical data by means
of a scaling theory deduced from the strong-noise RG \cite{VojtaHoyos15}.
Our paper is organized as follows. In Sec.\ \ref{sec:CP}, we define our model. Section
\ref{sec:Theory} is devoted to a summary of the strong-noise RG
and the resulting scaling theory. The Monte-Carlo simulations are presented in Sec.\
\ref{sec:MC}. We conclude in Sec.\ \ref{sec:Conclusions}.

\section{Contact process with temporal disorder}
\label{sec:CP}

The contact process \cite{HarrisTE74} is a prototypical lattice model featuring an
absorbing-state phase transition. It can be understood as a model for the spreading of an
epidemic. The contact process is defined on a
$d$-dimensional lattice which we assume to be hypercubic for simplicity. Each lattice
site can be in one of two states, healthy (inactive) or infected (active).
The time evolution of the contact process is a continuous-time Markov process during
which infected sites heal spontaneously at rate $\mu$ while healthy sites become
infected at rate $\lambda n/(2d)$. Here, $n$ is the number of infected neighbors
of the given healthy site. The long-time behavior of the contact process is determined
by the ratio between the infection rate $\lambda$ and the healing rate $\mu$.
If $\mu \gg \lambda$, the infection eventually dies out completely. The system ends up
in the absorbing state without any infected sites. This is the inactive phase.
In the opposite limit, $\lambda \gg \mu$, the density of infected sites remains nonzero
for all times. This is the active phase. In the clean case, when the rates $\lambda$ and
$\mu$ are uniform in space and independent of time, the phase transition between
the active and inactive phases is in the DP universality class.

We introduce temporal disorder, i.e., environmental noise, by making the infection and
healing rates time dependent. To be specific, we consider rates
\begin{equation}
\lambda(t) = \lambda_n ~,\quad
\mu(t) = \mu_n  \qquad (t_n < t < t_{n+1})
\label{eq:piecewise}
\end{equation}
that are piecewise constant over time intervals $\Delta t_n = t_{n+1}-t_n$.
The $\lambda_n$ and $\mu_n$ in different time intervals are statistically independent and drawn from probability
distributions $W_\lambda(\lambda)$ and $W_\mu(\mu)$.

\section{Theory}
\label{sec:Theory}

In this section, we summarize the strong-noise RG of Ref.\
\cite{VojtaHoyos15}, and we develop a scaling description of the phase transition.

\subsection{Mean-field theory}
\label{sec:MF_theory}

We start by considering the mean-field approximation of the temporarily disordered
contact process because its critical behavior can be found exactly.
The mean-field equation is obtained by assuming that the states of different
lattice sites are independent of each other. The time evolution of the density $\rho$
of active sites (for a single given realization of the temporal disorder)  is then
governed by the logistic evolution equation
\begin{equation}
\dot \rho(t) =  [\lambda(t)-\mu(t)] \rho(t) - \lambda(t) \rho^2(t)~.
\label{eq:MF-ODE}
\end{equation}
If the infection and healing rates $\lambda$ and $\mu$ are time-independent, this
differential equation can be solved in closed form. Employing this solution within
each time interval $(t_n,t_{n+1})$, we find a linear recurrence for the inverse
density of the given disorder realization,
\begin{equation}
\rho_{n+1}^{-1} = a_n  \rho_{n}^{-1} + c_n~.
\label{eq:mf-recurrence}
\end{equation}
Here, $\rho_n=\rho(t_n)$ is the density at the start of time interval $(t_n,t_{n+1})$.
The multipliers $a_n=\exp[(\mu_n-\lambda_n)\Delta t_n]$ implement the exponential growth
or decay due to the linear term in the evolution equation
(\ref{eq:MF-ODE}). The additive constants $c_n= (a_n-1)\lambda_n/(\mu_n-\lambda_n)$
limit the increase in $\rho$; they are only important for large $\rho$ and prevent
$\rho_n>1$.

The time evolution of the density therefore consists of a random sequence of spreading
(for $\lambda_n>\mu_n$) and decay (for $\lambda_n < \mu_n$) segments. This sequence
can either be mapped onto a random walk with a reflecting boundary
condition \cite{VojtaHoyos15}, or it can be analyzed by means of the strong-noise RG.

\begin{figure}
\includegraphics[width=7.5cm]{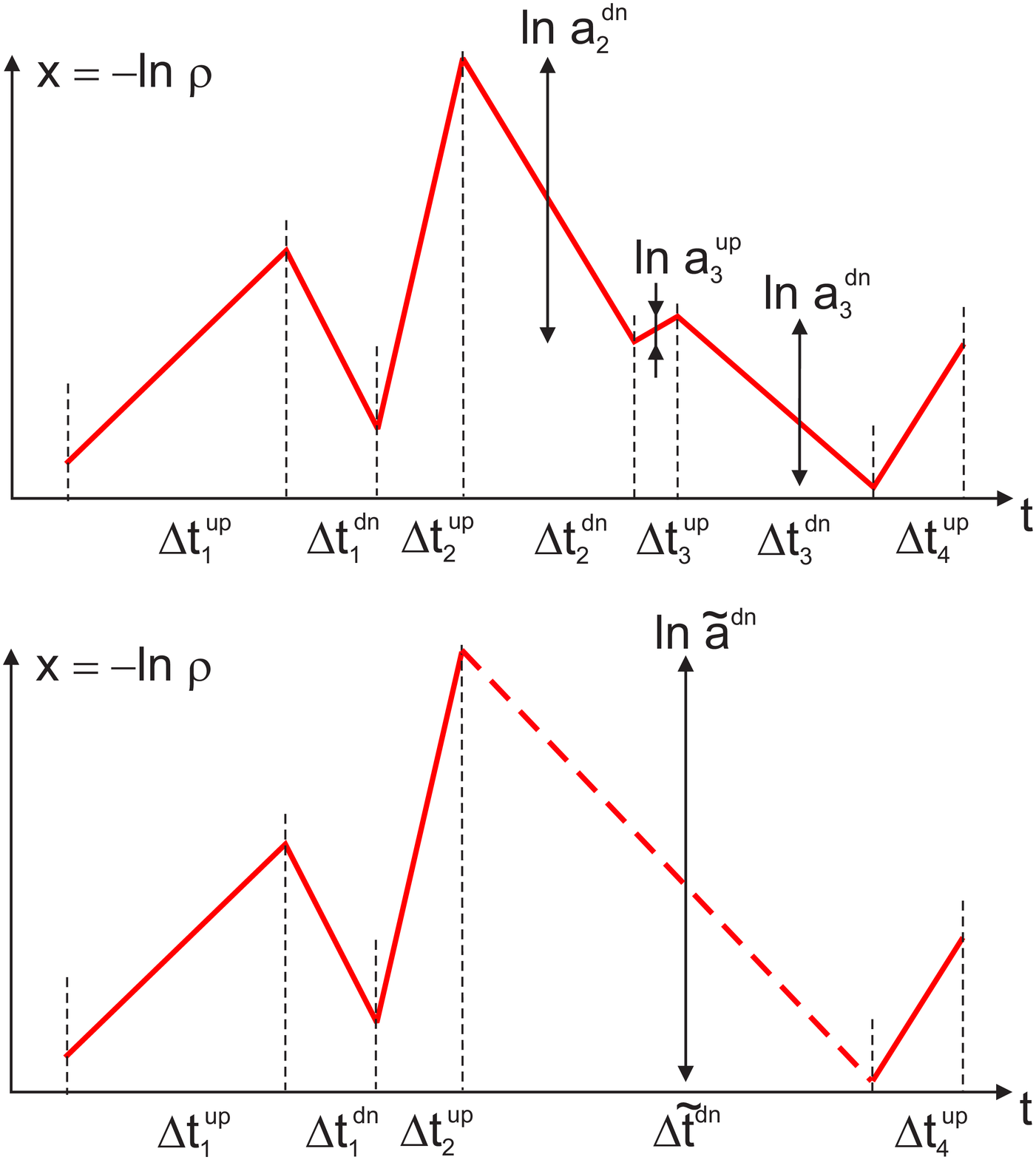}
\caption{(color online) Schematic of the strong-noise RG. The RG step replaces the time evolution in
the top panel by that in the bottom panel by eliminating the segment with the smallest
change of $x=-\ln\rho$, i.e, the segment with the smallest $|\ln a|$. Here, the segment
$\Delta t_3^{\rm up}$ with multiplier $a_3^{\rm up}$ is decimated by combining it with the segments
$\Delta t_2^{\rm dn}$ and $\Delta t_3^{\rm dn}$. This results in the renormalized time evolution
featuring the coarse-grained segment
$\Delta \tilde t^{\rm dn}$ with renormalized multiplier $\tilde a^{\rm dn}$, as given by Eqs.\  (\ref{eq:tilde_t_dn})
and (\ref{eq:tilde_a_dn}), respectively.}
\label{fig:RG_schematic}
\end{figure}

In the present paper, we focus on the RG because we will be able to adapt it
to finite-dimensional (non mean-field) systems later.
The  strong-noise RG, sketched in  Fig.\
\ref{fig:RG_schematic}, can be understood as the temporal analog of the
strong-disorder RG \cite{IgloiMonthus05} for spatially disordered systems.
We start by combining consecutive decay time intervals ($\mu_n>\lambda_n$) into a single
interval of length $\Delta t^\textrm{up}$. We also combine consecutive spreading intervals
($\mu_n<\lambda_n$) into a single interval of length $\Delta t^\textrm{dn}$. (Note that
``up'' and ``dn'' refer to the behavior of $x=-\ln\rho$.) The time evolution
is now a zig-zag curve of alternating spreading and decay segments.
In each segment, the inverse density evolves according to the recurrence
$\rho^{-1}(t+\Delta t) = a \rho^{-1}(t) + c$.
The multipliers of the spreading (down) segments
fulfill $a^\textrm{dn}<1$ while those of the decay (up) segments fulfill $a^\textrm{up} > 1$.

The strong-noise RG consists in iteratively decimating the weakest spreading and decay segments
which coarse-grains time. Specifically, each RG step eliminates the segment
for which the multiplier $a$ is closest to unity, i.e.,
the segment with the smallest $|\ln a|$, by combining it with the two neighboring segments,
as demonstrated in Fig.\ \ref{fig:RG_schematic}.
This defines the RG scale $\Gamma = \ln \Omega = \min(\ln a^\textrm{up}_i,-\ln a^\textrm{dn}_i)$.
The time evolution of $\rho^{-1}$ in the combined segment follows the same linear
recurrence $\rho^{-1}(t+\Delta \tilde t) = \tilde a \rho^{-1}(t) + \tilde c$
but with renormalized coefficients $\tilde a$ and $\tilde c$. If a spreading (down) segment
$a_i^\textrm{dn}$ is decimated, the renormalized multiplier reads
\begin{equation}
\tilde a^\textrm{up}=a_{i+1}^\textrm{up} a_i^\textrm{up} / \Omega~,
\label{eq:tilde_a_up}
\end{equation}
while the decimation of a decay (up) segment $a_i^\textrm{up}$ leads to
\begin{equation}
1/ \tilde a^\textrm{dn}=(1/a_{i}^\textrm{dn})\, (1/a_{i-1}^\textrm{dn}) / \Omega~,
\label{eq:tilde_a_dn}
\end{equation}
The renormalized additive constants $\tilde c$ are given by
$\tilde c^\textrm{up} = a_{i+1}^\textrm{up} a_i^\textrm{dn} c_{i}^\textrm{up} + a_{i+1}^\textrm{up} c_i^\textrm{dn} + c_{i+1}^\textrm{up}$ and
$\tilde c^\textrm{dn} = a_i^\textrm{dn} a_i^\textrm{up} c_{i-1}^\textrm{dn} + a_i^\textrm{dn} c_i^\textrm{up} + c_i^\textrm{dn}$
while the time intervals renormalize as
\begin{eqnarray}
\Delta\tilde t^\textrm{up} = \Delta t^\textrm{up}_{i} + \Delta t^\textrm{dn}_i + \Delta t^\textrm{up}_{i+1} ~,
\label{eq:tilde_t_up}
\\
\Delta\tilde t^\textrm{dn} = \Delta t^\textrm{dn}_{i-1} + \Delta t^\textrm{up}_i + \Delta t^\textrm{dn}_i ~.
\label{eq:tilde_t_dn}
\end{eqnarray}
All these RG recursion relations are exact, and the recursions for the multipliers $a^\textrm{up}$ and $a^\textrm{dn}$
are independent of the additive constants $c^\textrm{up}$ and $c^\textrm{dn}$. Upon iterating the RG
step, the probability distributions of the multipliers change, and their behavior in the limit $\Omega\to\infty$
determines the long-time physics.

The RG defined by the recursion relations (\ref{eq:tilde_a_up}) to (\ref{eq:tilde_t_dn}) is the temporal equivalent
of Fisher's RG for the spatially disordered transverse-field Ising chain \cite{Fisher92,*Fisher95} (with time taking
the place of position) and can be solved in the same way. The solution yields an exotic infinite-noise critical point at which the
distributions of $\ln a^\textrm{up}$ and $\ln a^\textrm{dn}$ become infinitely broad in the long-time limit. This leads
to enormous density fluctuations and an unusual logarithmic dependence of the life time on the system size.

The theory of this mean-field infinite-noise critical point was worked out in detail in Ref.\ \cite{VojtaHoyos15},
here we simply quote key results. The \emph{disorder-averaged} density at criticality decays as
$\rho_{\rm av} \sim t^{-\delta}$ with time $t$, the stationary density in the active phase
varies as $\rho_{\rm av} \sim |r|^\beta$ with distance $r$ from criticality and
the correlation time varies as $\xi_t \sim |r|^{-\nu_\parallel}$. The exponent values
\begin{equation}
\delta=1/2 ~,~\quad \beta =1 ~,\quad \nu_\parallel = 2~.
\label{eq:MF-exponents}
\end{equation}
differ from the ``clean'' mean-field exponents
$\delta=1,~ \beta =1,~ \nu_\parallel = 1$ \cite{Hinrichsen00}.
The probability distribution of the density, $P(x,t)$ with $x=-\ln \rho$, broadens without limit
with increasing time at criticality; this reflects the infinite-noise character of the critical point.
$P(x,t)$ obeys the single-parameter scaling form $P(x,t)= t^{-1/2}\, \Phi(x/t^{1/2})$.
Because of the infinite-noise character
of the critical point, this critical behavior is asymptotically exact.
Moreover, the lifetime $\tau_N$ of a finite-size sample in the active phase shows an anomalous power-law
dependence $\tau_N \sim N^{1/\kappa}$ on the sample volume (number of sites) $N$, in agreement with the
notion of a temporal Griffiths phase \cite{VBLM11}. The Griffiths exponent
$\kappa$ diverges at criticality, giving rise to the logarithmic dependence $\tau_N \sim \ln^2 N$.

\subsection{Finite dimensions}
\label{sec:Finite_d}

We now adapt the strong-noise RG to the case of the finite-dimensional (non mean-field) contact process.
Similar to the mean-field case, the time evolution is a sequence of density decay and spreading segments.
For strong temporal disorder, each individual segment is deep
in one of the two phases and far away from criticality. This suggests that one can neglect spatial fluctuations
and formulate the theory in terms of the time-dependent density $\rho(t)$ only.  We will return to the
validity of this approximation in Sec.\  \ref{sec:Conclusions}.

How does the time evolution of $\rho(t)$ in finite dimensions differ from the mean-field case?
During the decay segments, the density decreases exponentially just as in the mean-field case
because  each infected lattice site can heal independently.
In contrast, the behaviors of $\rho(t)$ during the spreading segments
in the mean-field and finite-dimensional cases are qualitatively different. In a finite-dimensional
system with short-range couplings, the infection cannot spread faster than ballistically.
In fact, in the active phase of the clean contact process, the spreading is known to be precisely
ballistic, i.e., the boundary between an active cluster and the surrounding inactive
area advances, on average, with constant speed (see, e.g., Sec.\ 6.3 of Ref.\ \cite{MarroDickman99}).

\begin{figure}
\includegraphics[width=\columnwidth]{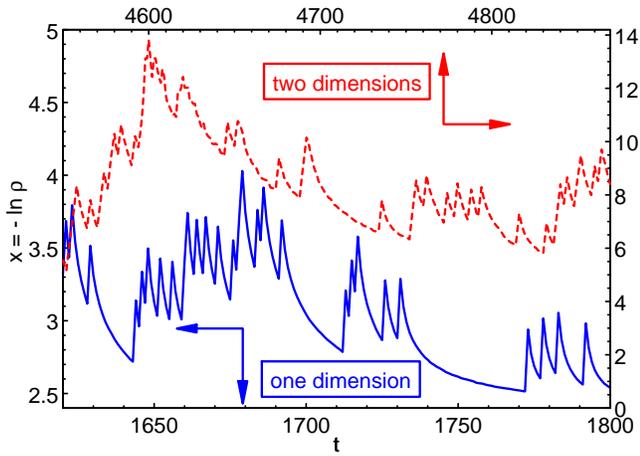}
\caption{(color online) Time evolution of the density of an individual noise realization
    at criticality,
    plotted as $x=-\ln \rho$ vs.\ $t$.
    The 1d data are for a system of $10^4$ sites
    using $\mu=1, \Delta t=1$ and binary distributed $\lambda$
     with $\lambda_h=8.1, \lambda_l=0.81$ and $p=0.8$.
    The 2d data are for $1000\times 1000$ sites
    using $\mu=1, \Delta t=2$ and binary distributed $\lambda$
     with $\lambda_h=3.65, \lambda_l=0.365$ and $p=0.8$
     (see Eq.\ (\ref{eq:W_binary}) for the meaning of the parameters).}
\label{fig:singleconf1d2d}
\end{figure}

In the case of strong temporal disorder, individual spreading segments are far away from
criticality. We therefore expect ballistic spreading during these segments. This implies
that the radius of an active cluster increases linearly with time, and its volume increases
as $t^d$. The total density of active sites during a spreading segment is proportional to
the total volume of all active clusters and therefore increases with time as
$\rho(t) \approx \rho_0 (1+bt)^d$  rather than an exponentially.
The qualitative difference between the exponential density decrease and the power-law increase can be
easily seen in the $\rho(t)$ curves of individual configurations of the temporal disorder
shown in Fig.\ \ref{fig:singleconf1d2d}.

We now modify the RG recursion relations to reflect the change in the spreading dynamics. When decimating a small
spreading (down) segment, the two neighboring exponential decay segments combine
multiplicatively just as in Eq.\ (\ref{eq:tilde_a_up}). In contrast, if a small decay (up) segment
is decimated, we need to combine its two neighboring ballistic spreading segments during which the
radii of active clusters increase linearly with time. The renormalized multiplier $\tilde a$ must reflect the
ballistic growth during entire renormalized time interval. For strong disorder,
it can be estimated as  $1/ \tilde a^\mathrm{dn}=(1+\tilde b\tilde \Delta t^{\rm dn})^d \approx (\tilde b\tilde \Delta t^{\rm dn})^d =
(b_i \Delta t_i^{\rm dn} + b_{i-1} \Delta t_{i-1}^{\rm dn})^d \approx
[(1/a_{i}^\mathrm{dn})^{1/d}+ (1/a_{i-1}^\mathrm{dn})^{1/d} ]^d$.
For the finite-dimensional contact process, we therefore arrive at the RG recursion
relations
\begin{eqnarray}
\tilde a^\textrm{up} &=& a_{i+1}^\textrm{up} a_i^\textrm{up} / \Omega~, \label{eq:tilde_a_up_fd} \\
(1/ \tilde a^\mathrm{dn})^{1/d} &=& (1/a_{i}^\mathrm{dn})^{1/d}+ (1/a_{i-1}^\mathrm{dn})^{1/d} - \Omega^{1/d}~.
\label{eq:tilde_a_dn_fd}
\end{eqnarray}
The last term in Eq.\ (\ref{eq:tilde_a_dn_fd}) contains the (subleading) contribution of
the decimated upward segment which we have added to make sure Eq.\ (\ref{eq:tilde_a_dn_fd}) is valid in the
atypical case $1/a_{i}^\mathrm{dn} = 1/a_{i-1}^\mathrm{dn} = \Omega$.
The time intervals renormalize according to (\ref{eq:tilde_t_up}) and (\ref{eq:tilde_t_dn})
as in the mean-field case.

The RG defined in Eqs.\ (\ref{eq:tilde_t_up}), (\ref{eq:tilde_t_dn}), (\ref{eq:tilde_a_up_fd}), and
(\ref{eq:tilde_a_dn_fd}) is formally equivalent
to the strong-disorder RG of spatially disordered quantum systems with super-Ohmic dissipation
\cite{VHMN11} or with long-range interactions \cite{JuhaszKovacsIgloi14}. To solve it, we
introduce  reduced variables $\Gamma=\ln\Omega$, $\beta=\ln a^\mathrm{up} -\Gamma$
and $\zeta=d[(\Omega a^\mathrm{dn})^{-1/d}-1]$. In terms of these variables, the flow equations for
the probability distributions ${\cal P}(\zeta;\Gamma)$  and ${\cal R}(\beta;\Gamma)$ read
\begin{equation}
\frac{\partial{\cal R}}{\partial\Gamma}  =  \frac{\partial{\cal R}}{\partial\beta}
                           +({\cal R}_0-{\cal P}_0){\cal R} +{\cal P}_0 \left({\cal R}\stackrel{\beta}{\otimes}{\cal R}\right) ~,
\label{eq:flow-R}
\end{equation}
\begin{equation}
\frac{\partial{\cal P}}{\partial\Gamma}  =  \left(1+\frac {\zeta} d \right)\frac{\partial{\cal P}}{\partial\zeta}
                           +\left({\cal P}_0-{\cal R}_0+\frac 1 d \right){\cal P} +{\cal R}_0\left({\cal P}\stackrel{\zeta}{\otimes}{\cal P}\right)~.
\label{eq:flow-P_finite_d}
\end{equation}
Here, ${\cal R}_0={\cal R}(0;\Gamma)$ and ${\cal P}_0={\cal P}(0;\Gamma)$, and the symbol ${\mathcal{P}}\stackrel{\zeta}{\otimes}{\mathcal{P}}=\int_{0}^{\zeta}{\mathcal{P}}(\zeta^{\prime}){\mathcal{P}}(\zeta-\zeta^{\prime}){\mathrm{d}}\zeta^{\prime}$
denotes the convolution.

\begin{figure}
\begin{center}
\includegraphics[width=7.5cm,clip]{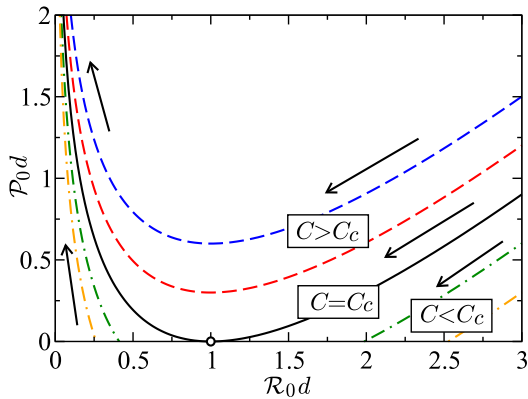}
\end{center}
\caption{(color online) Schematic of the RG flow in the ${\cal R}_0-{\cal P}_0$ plane. For $C<C_c$, the
flow asymptotically approaches a line of stable fixed points that represent the active phase.
In the inactive phase, $C>C_c$, the flow is towards ${\cal R}_0 \rightarrow 0$, and ${\cal P}_0 \rightarrow \infty$.
For $C=C_c$, one flows into the critical fixed point at ${\cal R}_0^*=1/d$, and ${\cal P}_0^*=0$.}
\label{fig:KT}
\end{figure}

The complete solution of the flow equations is rather complicated, but physically relevant solutions can be
obtained using the exponential ansatz
\footnote{The exponential ansatz can be motivated by the fact that its functional form is invariant under the convolution operation in
 Eqs.\  (\ref{eq:flow-R}) and (\ref{eq:flow-P_finite_d}). For the flow equations arising in the mean-field case, Fisher \cite{Fisher92,*Fisher95}
 showed analytically, that the fixed point solutions must have this form unless the bare disorder distributions are highly singular.
 In the case of Eqs.\ (\ref{eq:flow-R}) and (\ref{eq:flow-P_finite_d}), the same has been shown by numerically iterating the
 RG recursion relations \cite{JuhaszKovacsIgloi14}.}
\begin{equation}
{\cal R}(\beta;\Gamma)={\cal R}_0 e^{-{\cal R}_0 \beta} ~, \quad
{\cal P}(\zeta;\Gamma)={\cal P}_0 e^{-{\cal P}_0 \zeta}~.
\label{eq:ansatz}
\end{equation}
When we insert this ansatz into the flow equations  (\ref{eq:flow-R}) and (\ref{eq:flow-P_finite_d}),
we obtain the corresponding flow equations for the parameters ${\cal R}_0$ and ${\cal P}_0$,
\begin{equation}
d{\cal R}_0 / d\Gamma =-{\cal R}_0 {\cal P}_0~, \quad d{\cal P}_0 / d\Gamma =(1/d-{\cal R}_0){\cal P}_0
\label{eq:R0P0flow_finite_d}
\end{equation}
which take the well-known Kosterlitz-Thouless form \cite{KosterlitzThouless73}.
Let us discuss the fixed points of these flow equations and their properties. There is a line of fixed points at
${\cal P}_0^\ast=0,{\cal R}_0^\ast$ arbitrary. They are stable for ${\cal R}_0>1/d$ but unstable
for ${\cal R}_0<1/d$. The full RG trajectories in the ${\cal R}_0-{\cal P}_0$ plane can be obtained by combining
equations (\ref{eq:R0P0flow_finite_d}) to eliminate $\Gamma$. This yields
$d{\cal P}_0/d{\cal R}_0 =1-1/(d{\cal R}_0)$ with solution ${\cal P}_0 = {\cal R}_0 -(\ln {\cal R}_0) /d +C$
where $C$ is an integration constant. These Kosterlitz-Thouless type trajectories are sketched in Fig.\ \ref{fig:KT}.

Depending on the value of $C$, three regimes need to be distinguished:
(i) If $C>C_c=-(1+\ln d)/d$, the RG flow is towards ${\cal R}_0=0$ and ${\cal P}_0=\infty$.
This means that the upward multipliers $a^\textrm{up}$ become large and broadly distributed while the downward multipliers saturate at
$a^\textrm{dn}\approx 1/\Omega$. This is the inactive phase. (ii) For $C<C_c$, the flow is towards the line of stable
fixed points  ${\cal P}_0^\ast=0,{\cal R}_0^\ast>1/d$. Here, the downward multipliers
$1/a^\textrm{dn}$ become large and broadly distributed. This is the active phase. (iii) The critical point
corresponds to $C=C_c$ for which the flow approaches the endpoint ${\cal P}_0^\ast=0,{\cal R}_0^\ast=1/d$
of the line of stable fixed points.
To find the dependence of the renormalized time intervals on the RG scale $\Gamma$, we notice that every decimation
reduces the number $n(\Gamma)$ of up-down interval pairs by one. $n(\Gamma)$ thus fulfills the equation
\begin{equation}
dn / d\Gamma = -({\cal R}_0+{\cal P}_0)\, n~.
\label{eq:n(Gamma)}
\end{equation}
Expanding the RG flow equations (\ref{eq:R0P0flow_finite_d})
about the fixed points, we find the following long-time behavior in the active phase and at criticality:
$n(\Gamma) \sim \exp(-\Gamma {\cal R}_0^\ast)= \Omega^{-{\cal R}_0^\ast}$.
 The typical length of a renormalized time interval pair thus behaves as
$\Delta \tilde t \sim \Omega^{{\cal R}_0^\ast}$. At the critical point, this means
$\Delta \tilde t \sim \Omega^{1/d}$ because ${\cal R}_0^\ast=1/d$.
In the inactive phase, $\Delta \tilde t$ increases exponentially,
$\ln \Delta\tilde t \sim \Omega^{1/d}$.

Many physical results can be obtained by analyzing the RG. A central quantity is
the probability distribution $P(x)$ of the logarithm of the density, $x=-\ln\rho$.
Its width $\Delta x$ at criticality is determined by the typical value of $\ln a^\textrm{up}$.
As ${\cal R}_0^\ast=1/d$, we obtain $\Delta x \sim \ln a^\textrm{up} \approx \Gamma=\ln \Omega \sim \ln t$.
The distribution $P(x)$ thus broadens without limit, in agreement with the notion of ``infinite-noise criticality"
(and similar to the mean-field case discussed in Sec.\ \ref{sec:MF_theory}).
The behavior of the average density $\rho_\textrm{av} = \langle\rho \rangle$ can be found using the scaling ansatz $P(x)= \Phi(x/\ln t)/\ln t$
with a time-independent function $\Phi$. Integrating over $P(x)$, this gives $\rho_\textrm{av} \sim \langle \exp(-x) \rangle \sim (\ln t)^{-\bar\delta}$
with $\bar\delta =1$ (the overbar indicates the logarithmic rather than power-law time dependence).
In contrast, the typical density $\rho_\textrm{typ} \sim \exp(-\langle x \rangle)$ decays as a power of $t$.

The correlation time $\xi_t$ can be determined as the time at which the off-critical solution of (\ref{eq:R0P0flow_finite_d})
deviates appreciably from the critical one. As expected from a Kosterlitz-Thouless flow, this yields an exponential dependence,
$\ln \xi_t \sim |r|^{-\bar\nu_\parallel}$ with $\bar\nu_\parallel=1/2$. Here $r=C-C_c$ measures the distance
from criticality. In the active phase, the density reached at time $\xi_t$
scales as the stationary density, $\langle \rho_{st} \rangle \sim 1/\ln \xi_t \sim |r|^{\beta}$ with order parameter exponent
$\beta=1/2$.

The RG also allows us to calculate the life time $\tau_N$ of a finite-size sample of $N$ sites.
To find $\tau_N$, we follow the RG until the typical upward multiplier reaches $a^\textrm{up}=N$.
The corresponding renormalized time interval $\Delta \tilde t$ on this RG scale is the life time $\tau_N$
because it is the typical time for a decay segment in which the density decreases by a factor $1/N$.
From the solutions of the flow equations, we find
$\tau_N \sim N^{{\cal R}_0^\ast}$
in the active phase and at criticality.
The life time thus increases as a power law
\begin{equation}
\tau_N \sim N^{1/\kappa}
\label{eq:temp_Griffiths}
\end{equation}
rather than exponentially with $N$
which is a manifestation of temporal Griffiths singularities \cite{VBLM11,VojtaHoyos15}. The Griffiths exponent
$\kappa$ does not diverge at criticality but saturates at $\kappa_c=d$.
The relation  $\tau_N \sim N^{1/d} \sim L$ at the critical point implies a dynamical exponent of $z=1$.
In the inactive phase, the life time only increases logarithmically, $\tau_N \sim \ln N$.

The RG thus directly gives the critical exponents
\begin{equation}
\bar\delta=1 ~,~\quad \beta =1/2 ~,\quad \bar\nu_\parallel=1/2~,\quad z=1~.
\label{eq:finite-d-exponents}
\end{equation}
Other exponents such as $\bar \nu_\perp=\bar \nu_\parallel/z=1/2$ can be found from scaling relations.
Note that the usual correlation length and time exponents $\nu_\perp$ and $\nu_\parallel$
are formally infinite because correlation length and time depend exponentially on $r$.
Analogously, the usual density decay exponent $\delta$ vanishes.

\subsection{Heuristic scaling theory}
\label{sec:scaling}

The RG of Sec.\ \ref{sec:Finite_d} is formulated in terms of the density. It can therefore
be used directly to analyze experiments and simulations in macroscopic systems at finite densities.
In Monte Carlo simulations, this includes the usual density decays runs that start from a
fully active lattice. However, it cannot be used directly to analyze spreading experiments
or simulations such as Monte Carlo runs that start from a single active site (because the RG
does not contain the notion of an individual cluster).

We therefore formulate a heuristic scaling theory that is based on the RG results but can be
generalized to spreading experiments. The explicit RG results of Sec.\ \ref{sec:Finite_d} suggest
the scaling form
\begin{equation}
\rho_\textrm{av} (r,t,L) = (\ln b)^{-\beta/\bar\nu_\perp} \rho_\textrm{av}(r (\ln b)^{1/\bar\nu_\perp}, tb^{-z}, Lb^{-1})
\label{eq:rho_scaling}
\end{equation}
with exponents $\beta=1/2$, $\bar\nu_\perp=1/2$, and $z=1$
for the average density $\rho_\textrm{av}$ as function of time $t$, system size $L$ and the distance $r$ from criticality.
Here, $b$ is an arbitrary length scale factor. The time reversal symmetry of DP
\cite{GrassbergerdelaTorre79} still holds in the presence of uncorrelated
temporal disorder, as is demonstrated in Appendix \ref{sec:Appendix_A}.
The (average) survival probability in a spreading experiment therefore has the same scaling form as the
density,
\begin{equation}
P_s (r,t,L) = (\ln b)^{-\beta/\bar\nu_\perp} P_s(r (\ln b)^{1/\bar\nu_\perp}, tb^{-z}, Lb^{-1})~.
\label{eq:Ps_scaling}
\end{equation}
The RG as well as the scaling forms (\ref{eq:rho_scaling}) and (\ref{eq:Ps_scaling}) imply that the
critical system behaves as a system in the active phase, apart from logarithmic corrections.
(The critical fixed point is the end point of a line of fixed points that describe the active phase.)
We therefore expect the number $N_s$ of sites in the active cloud and its radius $R$
in a spreading experiment to behave analogously.
This suggests ballistic spreading with logarithmic corrections and yields the scaling forms
\begin{eqnarray}
N_s (r,t,L) &=& b^d(\ln b)^{-y_N} N_s(r (\ln b)^{1/\bar\nu_\perp}, tb^{-z}, Lb^{-1})~,~
\label{eq:Ns_scaling}\\
R (r,t,L) &=& b(\ln b)^{-y_R} R(r (\ln b)^{1/\bar\nu_\perp}, tb^{-z}, Lb^{-1})~.
\label{eq:R_scaling}
\end{eqnarray}
Here, $y_N$ and $y_R$ are the (yet unknown) exponents that govern the logarithmic corrections.
They are not independent of each other because $N_s \sim P_s \rho R^d$ which gives
$y_N = 2\beta/\bar\nu_\perp  + d y_R$.

Setting $L=\infty$, $r=0$, and $b=t^{1/z}=t$ in the scaling forms (\ref{eq:rho_scaling}) to (\ref{eq:R_scaling})
gives the time dependencies of the observables at criticality. We find
\begin{eqnarray}
\rho_\textrm{av}(t) &\sim& (\ln t)^{-\bar\delta} ~\quad\qquad\textrm{with} ~  \bar\delta=\beta/\bar \nu_\parallel= 1~, \label{eq:rho(t)}\\
P_s(t) &\sim& (\ln t)^{-\bar\delta} ~\quad\qquad\textrm{with} ~  \bar\delta=\beta/\bar \nu_\parallel= 1~, \label{eq:Ps(t)}\\
R(t) &\sim& t^{1/z} (\ln t)^{-y_R}  \,\quad\textrm{with}  ~ z=1 ~, \label{eq:R(t)}\\
N_s(t) &\sim& t^\Theta (\ln t)^{-y_N} ~~\,\quad\textrm{with} ~  \Theta=d/z = d~. \label{eq:Ns(t)}
\end{eqnarray}

\section{Monte Carlo simulations}
\label{sec:MC}

\subsection{Overview}
\label{sec:MC_overview}

In this section, we report the results of large-scale Monte Carlo simulations of the temporally disordered contact
process in one and two space dimensions.

Our numerical implementation of the contact process is an adaption to the case of temporal disorder
of the method proposed by Dickman \cite{Dickman99}. The simulation begins at time $t=0$ from
some configuration of active  and inactive lattice sites and consists of a sequence of events.
In each event an active site is randomly chosen from a list of all $N_a$ active sites. This site then
either infects a neighbor with probability $\lambda(t)/[1+ \lambda(t)]$ or it heals with probability $1/[1+ \lambda(t)]$.
For infection, one of the neighboring sites is chosen at random. The infection succeeds if this neighbor
is inactive. After the event, the time is incremented by $1/N_a$.

Temporal disorder is introduced by making the infection probability a piecewise constant function of time, $\lambda(t)= \lambda_n$
for $t_n < t <t_{n+1}$ with $t_n = n \Delta t$. Each $\lambda_n$ is independently drawn from the
binary probability distribution
\begin{equation}
W_\lambda(\lambda) = p \delta (\lambda-\lambda_h) +(1-p) \delta(\lambda-\lambda_l)
\label{eq:W_binary}
\end{equation}
Here, $p$ is the probability of having the\
higher infection rate $\lambda_h$ while $(1-p)$ is the probability for the lower infection
rate $\lambda_l$.
All results are averaged over many disorder realizations.  Note that in our implementation of the contact
process both the infection probability and the healing probability vary with time such that their
sum is constant and equal to unity.

Employing this method, we carried out two types of simulation runs. (i) Density decay simulations
start from a fully active lattice and monitor the time evolution of the density $\rho(t)$ of active sites.
(ii) Spreading simulations start from a single active site in an otherwise inactive lattice. Here,
we compute the survival probability $P_s(t)$ of the epidemic as well as the average number of sites $N_s(t)$
in the active cloud and its (mean-square) radius $R(t)$. For the spreading runs, the system size is chosen much bigger
than the largest active cloud, eliminating finite-size effects.

\subsection{One space dimension}
\label{sec:MC_1d}

We first consider a system with strong temporal disorder. In this case, we expect the infinite-noise physics
predicted by the RG to be visible already at short times. Specifically, we use
piecewise constant infection rates drawn from the distribution
$W_\lambda(\lambda)=p\,\delta(\lambda-\lambda_h)+(1-p)\,\delta(\lambda-\lambda_h/20)$
with probability $p=0.8$ and a time interval $\Delta t =6$. The transition is tuned by varying $\lambda_h$.

\begin{figure}
\includegraphics[width=8.5cm]{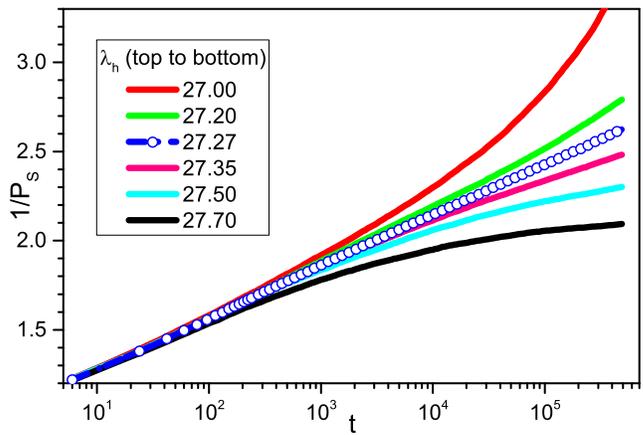}
\caption{(color online) Inverse survival probability $1/P_s$ vs.\ time $t$ for several values of the infection rate $\lambda_h$.
The data are averages over $30,000$ to $50,000$ disorder realizations, with one simulation run per realization. The
statistical error of the data is about one symbol size.}
\label{fig:Ps_t_1d}
\end{figure}

\begin{figure}
\includegraphics[width=8.5cm]{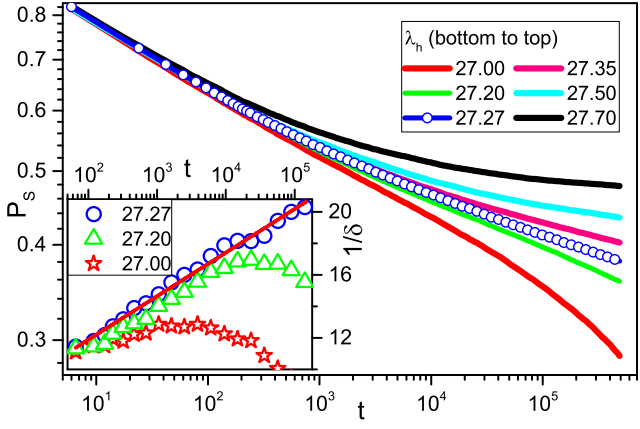}
\caption{(color online) Double logarithmic plot of the data of Fig.\ \ref{fig:Ps_t_1d}.
Inset: Inverse effective exponent $1/\delta_{\rm eff}$ vs.\ time $t$.}
\label{fig:Ps_t_1d_lnln}
\end{figure}

Figure \ref{fig:Ps_t_1d} shows the survival probability $P_s(t)$ of spreading runs as a function of time $t$,
plotted such that the predicted logarithmic decay (\ref{eq:Ps(t)}) corresponds to a straight line.
This yields a critical infection rate of $\lambda_h \approx 27.27(4)$  where the number in parentheses is an estimate
for the error of the last digit. At this infection rate, the data follow the prediction (\ref{eq:Ps(t)}) over almost four orders
of magnitude in time. The data for higher and lower $\lambda_h$ curve away from the straight line as expected.
To test whether these data could also be interpreted in terms of conventional power-law scaling,
we replot them in Fig.\ \ref{fig:Ps_t_1d_lnln} in a double logarithmic fashion (such that power laws correspond to
straight lines).
The figure demonstrates that
the critical survival probability cannot be described by a power law over any appreciable time interval.
To further confirm this observation, we calculate an effective (running) value for the conventional
decay exponent $\delta$ via $\delta_{\rm eff} (t)= - d\ln P_s(t)/ d \ln t$. If the critical behavior was of power-law type,
$\delta_{\rm eff}$ should approach the true asymptotic exponent $\delta$ with increasing time.
Instead, the data presented in the inset of Fig.\  \ref{fig:Ps_t_1d_lnln} show that $\delta_{\rm eff}$
decays like $1/\ln t$, exactly as expected from Eq.\ (\ref{eq:Ps(t)}). Slightly subcritical $\delta_{\rm eff}$
curves first follow the critical curve and then turn around, with $\delta_{\rm eff}$ now increasing
with time rather than saturating. The data are therefore incompatible with power-law scaling.

\begin{figure}
\includegraphics[width=8.5cm]{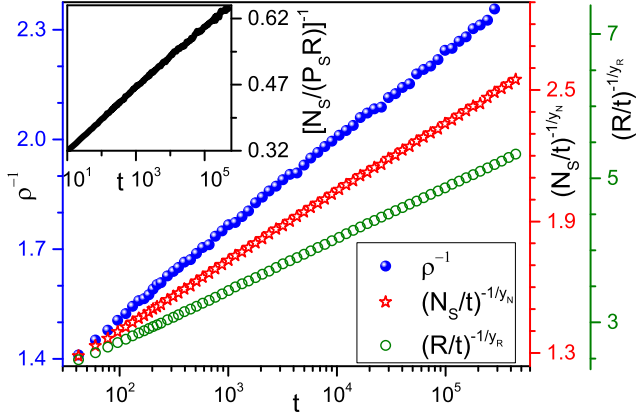}
\caption{(color online) Number of sites in the active cloud and its radius at criticality, $\lambda_h=27.27$, plotted as
$(N_s/t)^{-1/y_N}$ and $(R/t)^{-1/y_R}$ vs.\ time $t$ with $y_N=3.6$ and $y_R=1.7$. The data are averages over 50,000 disorder realizations
with one run per configuration, their errors are about one symbol size.
Also shown is the inverse average density for decay runs at criticality (system of 200,000 sites, 50,000 disorder configurations).
Inset: Density $N_s/(P_s R)$ of active sites inside an active cloud in a critical spreading run.}
\label{fig:rhoNsR_t_1d}
\end{figure}

The number $N_s$ of sites in the active cloud and its radius $R$ at criticality are shown in Fig.\ \ref{fig:rhoNsR_t_1d}.
To test the predictions (\ref{eq:R(t)}) and (\ref{eq:Ns(t)}) of the scaling theory, viz.\ ballistic growth with logarithmic corrections, we divide out the
ballistic behavior $N_s\sim R \sim t$. We then plot $(N_s/t)^{-1/y_N}$ and $(R/t)^{-1/y_R}$ vs.\ $\ln t$ and vary the exponents $y_N$ and $y_R$ until the curves
are straight lines. The data follow the predicted behavior over more than three orders of magnitude in time which confirms $z=1$.
Moreover, the resulting exponent values, $y_N=3.6(4)$ and $y_R=1.7(3)$, fulfill the relation $y_N= 2 \beta/\bar\nu_\perp + d y_R$
(using the predicted value $\beta/\bar\nu_\perp=1$).

In addition to the spreading runs, we have also performed density decay simulations. Figure \ref{fig:rhoNsR_t_1d} demonstrates
that time dependence of the average density $\rho_\textrm{av}$ at the critical infection rate follows the predicted logarithmic behavior (\ref{eq:rho(t)}).
For comparison, the density of active sites inside a (surviving) active cloud in a spreading simulation can be found
from the combination $N_s/(P_s R)$. The inset of Fig.\ \ref{fig:rhoNsR_t_1d} shows that this density behaves as $1/\ln(t)$, just as the
density $\rho_\textrm{av}$ of a decay simulation.

\begin{figure}
\includegraphics[width=8.5cm]{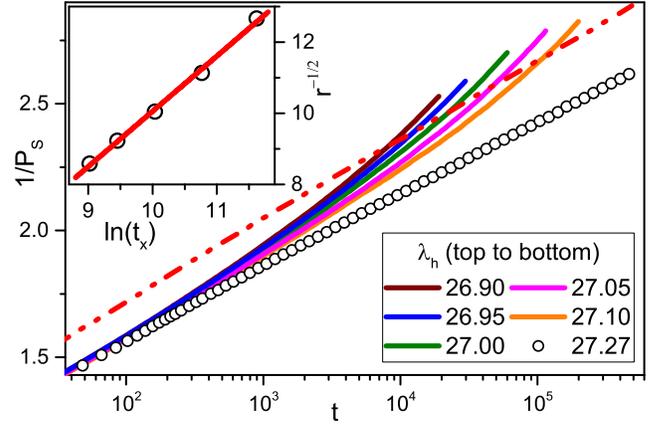}
\caption{(color online) Inverse survival probability $1/P_s$ vs.\ time $t$ for infection rates $\lambda_h$ at and below the critical rate $\lambda_c=27.27$.
The data are averages over $5\times10^{4}$ to $10^6$ disorder realizations with one run per configuration,
their errors are about a symbol size.
The dash-dotted line shows $1.1/P_s$ for $\lambda=\lambda_c$.
Inset: Distance from criticality $r=(\lambda_c-\lambda)/\lambda_c$ vs.\ crossing time $t_x$, plotted as $r^{-1/2}$ vs.\ $\ln(t_x)$.}
\label{fig:Ps_offcritical_1d}
\end{figure}

In order to extract the complete critical behavior from the simulations, we also analyze the off-critical survival probability.
Figure \ref{fig:Ps_offcritical_1d} shows $1/P_s$ as a function of $t$ for several infection rates slightly below the critical rate.
The crossings of the off-critical curves with the line representing $1.1/P_s$ for $\lambda=\lambda_c$ define the crossover times
$t_x (\lambda)$. According to (\ref{eq:Ps_scaling}), these crossover times should depend on the distance $r$ from criticality via
$\ln(t_x) \sim r^{-1/2}$. The inset of Fig.\ \ref{fig:Ps_offcritical_1d} demonstrates that this relation is fulfilled with
reasonable accuracy.

In addition to the averages of $P_s, N_s$ and $R$, we have also studied the time evolution of their probability distributions
(w.r.t. the temporal disorder).
These simulations require a particularly high numerical effort
because we need to perform many runs for each individual disorder configuration to obtain reliable values for $P_s$, $N_s$, and $R$.
This limits the maximum simulation time.

\begin{figure}
\includegraphics[width=8.5cm]{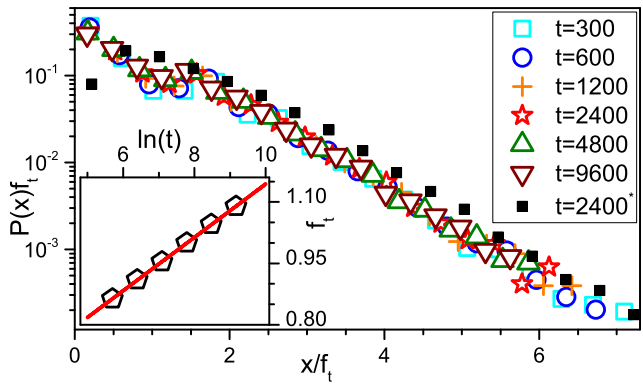}
\caption{(color online) Probability distribution $P(x,t)$ with $x=-\ln P_s$ at criticality for different times $t$.
The argument $x$ is scaled by a factor $f_t$ such that the curves at different times coincide.
The data are averages over $20,000$ disorder realizations with 1000 runs for
each configuration. For comparison, the solid black squares show $P(x,t)$ for a system
with a box disorder distribution (see text). Inset:  Scale factor $f_t$ vs.\ $\ln t$.}
\label{fig:Ps_distrib_1d}
\end{figure}

The probability distribution $P(x,t)$ of the logarithm of the survival probability,  $x=-\ln P_s$,
at criticality broadens without limit with increasing time $t$, in agreement with the notion of infinite-noise criticality.
If we rescale the width by $\ln t$, the distributions at different times $t$ all fall onto a single master curve.
This is demonstrated in Fig.\ \ref{fig:Ps_distrib_1d} which shows
a scaling plot of the distribution $P(x,t)$ at criticality.
The data for all considered times scale very well; and the scale factor $f_t$ depends linearly on $\ln t$.
This implies that the distribution fulfills single-parameter scaling; it has the scaling form
$P(x,t) = \Phi_P(x/\ln t)/\ln t$ with $\Phi_P$ being a time-independent scaling function.

Is the
scaling function $\Phi_P$ universal (i.e., independent of the disorder distribution)?
As the value of the survival probability depends not only on the late stages of the time evolution
which are governed by the universal RG fixed point, but also on the initial time steps which are controlled
by the bare disorder distribution, we do not expect the scaling function $\Phi_P$ to be universal.
In particular, the behavior close to $x=0$ (i.e., $P_s=1$) is dominated by atypical disorder realizations
that contain only large infection rates. However, we expect the functional form of the tail of the
distribution to be universal in the $t\to \infty$ limit because it is governed by the RG fixed point.
To test this numerically, we have performed a set of simulations with box-distributed disorder,
($\lambda$ uniformly distributed between $\lambda_h/10$ and $\lambda_h$),
rather than the binary disorder (\ref{eq:W_binary}). The resulting $P(x,t)$ at criticality ($\lambda_h=\lambda_c=10.135$)
is included in Fig.\ \ref{fig:Ps_distrib_1d} for one characteristic time. The plot shows that the distribution is indeed
nonuniversal close $x=0$, but the functional form of the tail agrees with the results from the
binary disorder.

\begin{figure}
\includegraphics[width=8.5cm]{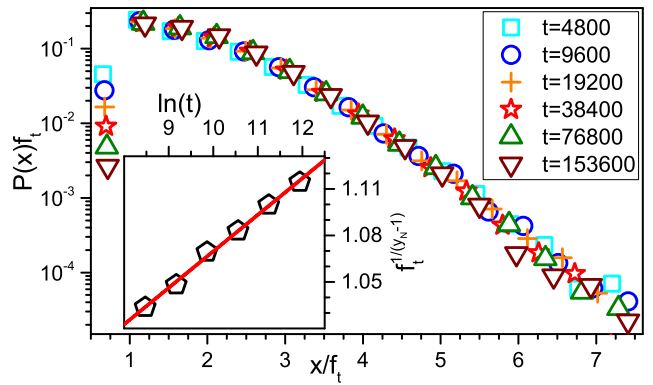}
\caption{(color online) Probability distribution $P(x,t)$ with $x=-\ln(N_{S}P_{S}^{-1}t^{-1})$ at criticality for different times $t$,
scaled such that the curves coincide. The data are averages over $500,000$ disorder realizations. Inset:  Scale factor $f_t^{1/(y_N-1)}$ vs.\ $\ln t$
with $y_N=3.7$.}
\label{fig:Ns_distrib_1d}
\end{figure}

The probability distribution of the average number $N_s$ of sites in the active cloud can be analyzed analogously. Specifically, we consider
$N_s/P_s$ (the number of active sites in a \emph{surviving} cloud), and we divide out the leading factor $t$ [see Eq.\ (\ref{eq:Ns(t)})] to
focus on the logarithmic corrections. Figure  \ref{fig:Ns_distrib_1d} presents a scaling plot of the distribution $P(x,t)$ of $x=-\ln(N_{s}P_{s}^{-1}t^{-1})$
at the critical infection rate and different values of the time $t$.
The data scale very well, and the scale factor varies as $\ln(t)^{y_N-1}$ as suggested by the combination of eqs.\   (\ref{eq:Ps(t)}) and (\ref{eq:Ns(t)}).
This implies that the distribution takes the scaling form
$P(x,t) = \tilde \Phi_N[x/(\ln t)^{y_N-1}]/(\ln t)^{y_N-1)}$ where $\Phi_N$ is another time-independent scaling function.

\begin{figure}
\includegraphics[width=8.5cm]{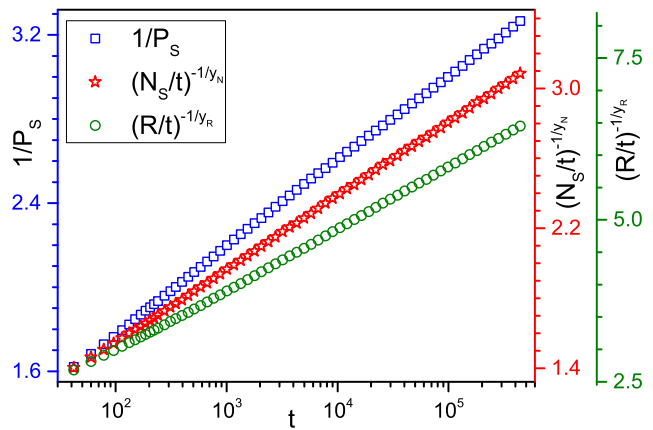}
\caption{(color online) $P_s^{-1}$,  $(N_s/t)^{-1/y_N}$, and $(R/t)^{-1/y_R}$ vs.\ time $t$ at criticality ($\lambda_h=20.49$) for
a system with a shorter disorder time interval $\Delta t =3$. The exponents are fixed at $y_N=3.6$ and $y_R=1.7$.
The data are averages over 100,000 disorder realizations with one run per configuration, leading to errors
of about one symbol size.}
\label{fig:Delt3_1d}
\end{figure}

All simulations reported so far confirm the strong-noise RG and the scaling theory
of Sec.\ \ref{sec:Theory}. How universal is this conclusion?
Fig.\ \ref{fig:Delt3_1d} presents the results of spreading runs for a system
with a shorter base time interval  $\Delta t$ of the piecewise constant disorder in $\lambda(t)$.
(The disorder distribution is $W_\lambda(\lambda)=p\,\delta(\lambda-\lambda_h)+(1-p)\,\delta(\lambda-\lambda_h/20)$
with probability $p=0.8$ and $\Delta t =3$.)
The shorter base time interval, viz. $\Delta t =3$ instead of 6,
reduces the probability for finding long (rare) time periods during which the infection rate does not change.
Figure \ref{fig:Delt3_1d} demonstrates that $P_s$, $R$, and $N_s$ at criticality nonetheless follow
the predictions (\ref{eq:Ps(t)}), (\ref{eq:R(t)}), and (\ref{eq:Ns(t)}) of the scaling theory with
the same exponents $y_R=1.7$ and $y_N=3.6$ as the earlier system.

\begin{figure}
\includegraphics[width=8.5cm]{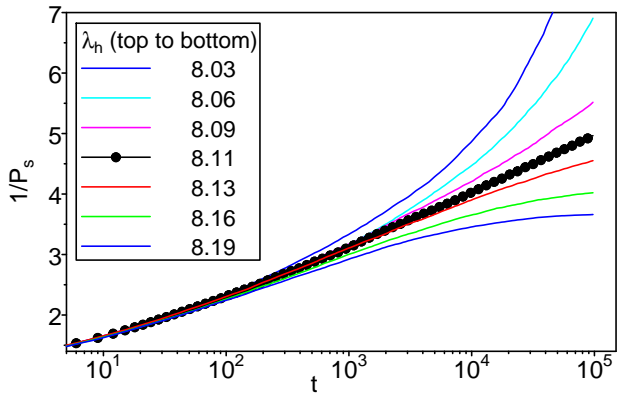}
\caption{(color online) Inverse survival probability $1/P_s$ vs.\ time $t$ for several values of the infection rate $\lambda_h$
in a system with weaker temporal disorder (see text).
The data are averages over $300$ disorder realizations, with 100 simulation runs per realization.
Note the crossover to logarithmic behavior at $t\approx 10^3$.}
\label{fig:Ps_t_weak_1d}
\end{figure}

Does our theory also hold for even weaker disorder? To address this question, we have carried out simulations for
several additional disorder distributions covering the range from moderate to weak disorder. In all cases, observables
at criticality display deviations from the power-law behavior expected at conventional critical points.
In particular, the average density of critical decay runs as well as the critical survival probability of critical
spreading runs decrease more slowly than a power law with time \footnote{For weak disorder, these deviations are subtle and only
visible in high-precision data.}.
However, the crossover from the clean critical point to the true asymptotic behavior is very slow, perhaps because
the violation of Kinzel's stability criterion $\nu_\parallel>2$ is not very strong. (The clean correlation time
exponent takes the value $\nu_\parallel \approx 1.73$ in one dimension \cite{Jensen99}.)
This is illustrated in Fig.\ \ref{fig:Ps_t_weak_1d} which presents the survival probability of spreading
simulations of a moderately disordered system having a distribution
$W_\lambda(\lambda)=p\,\delta(\lambda-\lambda_h)+(1-p)\,\delta(\lambda-\lambda_h/10)$ with probability $p=0.8$ and $\Delta t =1$.
The figure shows that the initial decay of the critical $P_s$ with time is faster than the logarithmic behavior (\ref{eq:Ps(t)}),
as indicated by the upward curvature of the data, but  after a crossover time $t_x \approx 1000$, the data settle on the
straight line expected from theory. Assuming a generalized logarithmic dependence, $P_s(t) \sim \ln^{-x}(t)$ with $x\ne 1$
does not improve the fit. We have also performed a local slope analysis assuming power-law scaling analogous to Fig.\ \ref{fig:Ps_t_1d_lnln}.
As in that case, the effective exponent $\delta_{\rm eff}$ approaches zero with increasing time, as expected from our theory.

For even weaker disorder, the crossover from the clean to the disordered critical behavior is even later which puts the asymptotic regime beyond the
range of our numerical capabilities.

\subsection{Two space dimensions}
\label{sec:MC_2d}

We again begin the discussion by considering a system with strong temporal disorder, characterized by the binary distribution
$W_\lambda(\lambda)=p\,\delta(\lambda-\lambda_h)+(1-p)\,\delta(\lambda-\lambda_h/10)$
with probability $p=0.8$ and a long base time interval of $\Delta t =6$.

\begin{figure}
\includegraphics[width=8.5cm]{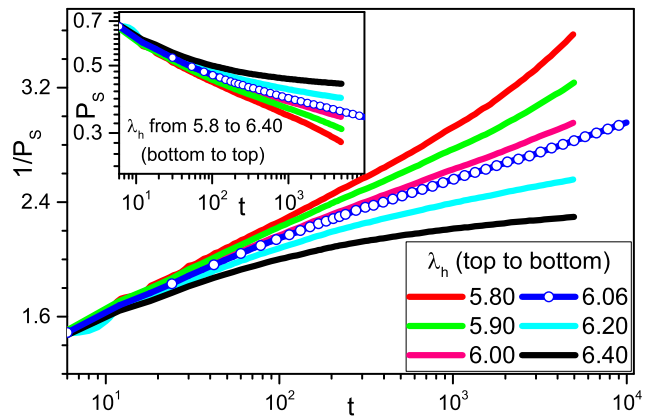}
\caption{(color online) Inverse survival probability $1/P_s$ vs.\ time $t$ in two space dimensions for several values of the infection rate $\lambda_h$.
The data are averages over $60,000$ to $120,000$ disorder realizations, with one simulation run per realization. The
statistical error of the data is about one symbol size. Inset: Double logarithmic plot of $P_s$ vs.\ $t$.}
\label{fig:Ps_t_2d}
\end{figure}

Figure \ref{fig:Ps_t_2d} presents the survival probability $P_s$ of spreading simulations as function of $t$.
The data at the critical infection rate $\lambda_h=6.06$ follow the predicted logarithmic behavior (\ref{eq:Ps(t)}) over about two
orders of magnitude in time, confirming the theory. To further support this conclusion, the inset of this figure presents a double
logarithmic plot of $P_s$ vs.\ $t$ which demonstrates that power laws do not describe the data over any appreciable time interval.

\begin{figure}
\includegraphics[width=8.5cm]{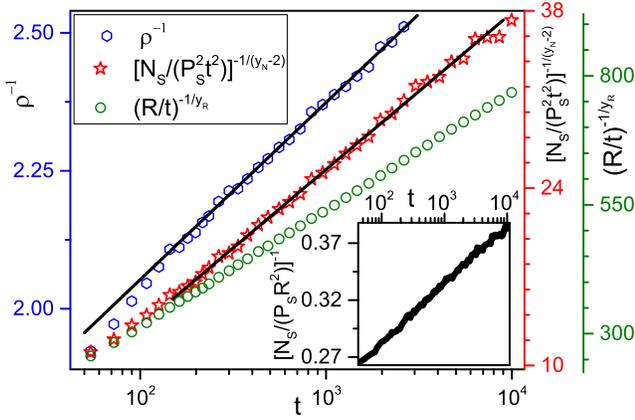}
\caption{(color online) Number of sites in the active cloud and its radius in two dimensions at criticality, $\lambda_h=6.06$, plotted as
$(N_s/t)^{-1/y_N}$ and $(R/t)^{-1/y_R}$ vs.\ time $t$ with $y_N=2.5(3)$ and $y_R=0.29(5)$. (60,000 disorder
realizations with one run per configuration, the errors are about one symbol size.)
Also shown is the inverse average density $\rho_\textrm{av}^{-1}$ for decay runs (system of $3200^2$ sites, 60,000 disorder configurations).
Inset: Density $N_s/(P_s R^2)$ of active sites inside an active cloud in a critical spreading run.}
\label{fig:rhoNsR_t_2d}
\end{figure}

The time dependencies of the number $N_s$ of sites in the active cloud and
its radius $R$ at criticality are shown in Fig.\ \ref{fig:rhoNsR_t_2d}.
To verify the predictions (\ref{eq:R(t)}) and (\ref{eq:Ns(t)}) of the scaling theory, we divide out the ballistic power laws
$R \sim t$ and $N_s \sim t^2$. We then plot $(R/t)^{-1/y_R}$ and $(N_s/t^2)^{-1/y_N}$ vs.\ $\ln t$ and vary the exponents
$y_N$ and $y_R$ until the curves are straight lines which gives $y_N=2.5(3)$ and $y_R=0.29(5)$. Figure \ref{fig:rhoNsR_t_2d}
thus confirms that $R$ and $N_s$ follow the scaling theory. The figure also shows the time dependence of the average density
$\rho_\textrm{av}$ of decay runs at criticality. It follows the predicted logarithmic behavior (\ref{eq:rho(t)}). The inset of Fig.\
\ref{fig:rhoNsR_t_2d} shows the average density of active sites inside a (surviving) active cloud in the spreading simulations,
as given by the combination $N_s/(P_s R^2)$. It follows the same logarithmic time dependence as the average density measured in
decay simulations.

\begin{figure}
\includegraphics[width=8.5cm]{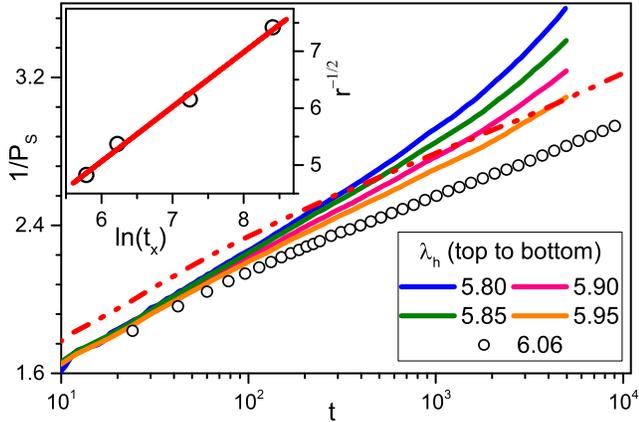}
\caption{(color online) Inverse survival probability $1/P_s$ vs.\ time $t$ for infection rates $\lambda_h$ at and below the critical rate $\lambda_c=6.06$.
(60,000 to 100,000 disorder realizations with one run per configuration,
giving statistical errors of about a symbol size.)
The dash-dotted line represents $1.09/P_s$ for $\lambda=\lambda_c$.
Inset: Distance from criticality $r=(\lambda_c-\lambda)/\lambda_c$ vs.\ crossing time $t_x$, plotted as $r^{-1/2}$ vs.\ $\ln(t_x)$.}
\label{fig:Ps_offcritical_2d}
\end{figure}

As in the one-dimensional case, we also analyze the off-critical behavior of the survival probability.
Figure \ref{fig:Ps_offcritical_2d} presents $1/P_s$ vs.\ $t$ for several infection rates slightly below the critical one.
The crossings of the off-critical curves with the line representing $1.09/P_s$ at criticality define the crossover times
$t_x (\lambda)$. These crossover times are predicted to depend on the distance $r$ from criticality via
$\ln(t_x) \sim r^{-1/2}$ [see Eq.\ (\ref{eq:Ps_scaling})]. The inset of Fig.\ \ref{fig:Ps_offcritical_2d} confirms
that this relation is fulfilled.

To test the universality of the critical behavior, we have also performed density decay simulations
of a system with disorder distribution $W_\lambda(\lambda)=p\,\delta(\lambda-\lambda_h)+(1-p)\,\delta(\lambda-\lambda_h/10)$,
with $p=0.8$ and a shorter base time interval $\Delta t=2$. These simulations were presented in Fig.\ 3 of Ref.\ \cite{VojtaHoyos15}
to illustrate the strong-disorder RG theory.
We found
that the average density at criticality ($\lambda_h=3.65$) decays logarithmically with time, as predicted in Eq.\
(\ref{eq:rho(t)}).
Analogously to Figs.\ \ref{fig:Ps_distrib_1d} and \ref{fig:Ns_distrib_1d}, the probability distribution $P(x,t)$ with $x=-\ln\rho$
at different times $t$ scales very well, and the scale factor depends linearly on $\ln t$, as expected.

\subsection{Temporal Griffiths phases}
\label{sec:Griffiths}

Vazquez et al.\ \cite{VBLM11} introduced the concept of a temporal Griffiths phase in a temporally
disordered system. The temporal Griffiths phase is the part of the active phase in which the life time $\tau_N$
of a finite-size sample shows an anomalous power-law dependence on the system size $N$ (as opposed to the exponential
dependence expected in the absence of temporal disorder).

Our strong-noise RG for the contact process with temporal disorder predicts such power-law
behavior, $\tau_N \sim N^{1/\kappa}$, see Eq.\ (\ref{eq:temp_Griffiths}). Moreover, it predicts
that the Griffiths exponent $\kappa$ increases monotonically as the critical point is
approached from the active side and saturates at the value $\kappa_c=d$ at criticality.

\begin{figure}
\includegraphics[width=8.5cm]{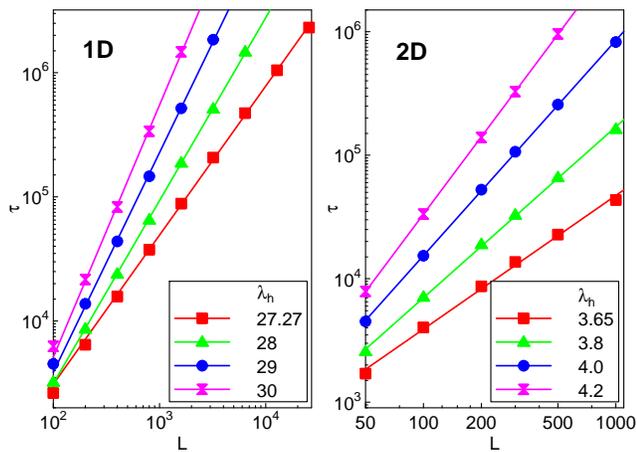}
\caption{(color online) Log-log plots of the average life time $\tau$ of samples of linear
size $L=N^{1/d}$ for several infection rates $\lambda_h$ at and above the critical point.
The solid lines are power law fits. The data are averages over 10,000 to 20,000 disorder
configurations.
Left: One-dimensional system with critical infection rate $\lambda_h\approx 27.27$.
Right: Two-dimensional system with critical infection rate $\lambda_h \approx 3.65$. }
\label{fig:tempgriffiths1d2d}
\end{figure}

To test these predictions, we have performed density decay simulations (starting from a fully active lattice)
on finite-size samples. We have
measured the average life time $\tau_N$, i.e., the average of the time at which a sample of $N$ sites reaches the
absorbing state. Figure \ref{fig:tempgriffiths1d2d} shows the results for  both one and two-dimensional systems.
In the one-dimensional case, we use the same parameters as in the main part of
Sec.\ \ref{sec:MC_1d}, i.e., piecewise constant infection rates having the distribution
$W_\lambda(\lambda)=p\,\delta(\lambda-\lambda_h)+(1-p)\,\delta(\lambda-\lambda_h/20)$
with $p=0.8$ and a time interval $\Delta t =6$. For these parameters, the critical point is
located at $\lambda_h=\lambda_c\approx 27.27$. The left panel of  Fig.\ \ref{fig:tempgriffiths1d2d}
demonstrates that the life time indeed follows the power law $\tau \sim N^{1/\kappa} = L^{d/\kappa}$
both at criticality and slightly in the active phase. The Griffiths exponent $\kappa$ is non-universal
and decreases with increasing $\lambda_h$, as predicted.
The right panel  of  Fig.\ \ref{fig:tempgriffiths1d2d} shows analogous behavior in the
two-dimensional case. Here we use the disorder distribution $W_\lambda(\lambda)=p\,\delta(\lambda-\lambda_h)+(1-p)\,\delta(\lambda-\lambda_h/10)$,
with $p=0.8$ and a time interval $\Delta t=2$ which yields a critical infection rate of  $\lambda_h=\lambda_c\approx 3.65$.

Straight power-law fits of the life time at criticality to $\tau \sim N^{1/\kappa} = L^{d/\kappa}$
 give exponents $\kappa_c \approx 0.85$ and 1.9 in one and two dimensions, respectively. These values
 agree reasonably well with the RG prediction of $\kappa_c=d$. Moreover, the data at criticality in Fig.\
\ref{fig:tempgriffiths1d2d} show a slight downward curvature which suggests corrections to the leading
power-law behavior. Indeed, the data can be fitted very well to the predicted power law with exponent
$\kappa_c=d$ if a correction-to-scaling term is included.

\section{Conclusions}
\label{sec:Conclusions}

In summary, we have performed large-scale Monte Carlo simulations of the contact process in the presence of temporal disorder,
i.e., external environmental noise, in one and two space dimensions. The purpose of the simulations was to test the recently developed
real-time strong-noise RG theory \cite{VojtaHoyos15} for temporally disordered systems. This theory
predicts an exotic ``infinite-noise'' critical point which can be understood as the temporal counterpart of the infinite-randomness
critical points found in the spatially disordered contact process and other systems. According to the RG theory,
the width of the density distribution at criticality diverges in
the long-time limit, even on a logarithmic scale, and the dynamics of the average density as well as the survival probability
become logarithmically slow.

The strong-noise RG for the finite-dimensional contact process takes spatial fluctuations into account only approximately (by treating the density increase
during the spreading segments as ballistic). We expect this to be a good approximation for strong temporal disorder because in this case
individual spreading and decay segments are far away from criticality. As the temporal disorder increases under the RG,
this condition seems to be asymptotically fulfilled. Furthermore, the clean critical point violates Kinzel's stability criterion
$\nu_\parallel >2$ in all dimensions which implies that even weak temporal disorder is a relevant perturbation and grows under coarse graining.
These arguments suggest that the RG theory gives the correct critical behavior for any (nonzero) bare disorder
strength. However, a rigorous proof that the fixed point found by the strong-noise RG is stable will require
a proper analysis of spatial fluctuations in addition to the temporal ones. This is beyond the scope of the present
theory
\footnote{If the only relevant effect of spatial fluctuations was to change the time dependence of the density during spreading segments
from ballistic to a weaker power law, $\rho(t) \approx \rho_0 (1+b t)^{d'}$ with $d'<d$, our theory would change very little.
The recursion for $1/ \tilde a^\mathrm{dn}$ would take the additive form (\ref{eq:tilde_a_dn_fd})
with $d$ replaced by $d'$. The resulting critical behavior \cite{VHMN11} would again be of Kosterlitz-Thouless form
with the critical exponents given by (\ref{eq:finite-d-exponents}) except the dynamical exponent which would take
the value $z=d/d'$.}.

To test the theoretical predictions, we have simulated systems with both strong and weak (bare) temporal disorder.
Our simulations for strongly disordered systems fully confirm the results of the RG and the related
heuristic scaling theory in both one and two space dimensions; conventional power-law scaling can be excluded.
For weak and moderately strong disorder, the crossover
from the clean critical fixed point to the true asymptotic behavior is very slow, in particular in one dimension where the violation
of Kinzel's stability criterion $\nu_\parallel > 2$ is not very pronounced. (The clean correlation time exponent takes the values
$\nu_\parallel$ approx 1.73 in one dimension and 1.29 in two dimensions \cite{Dickman99}.) For moderately strong disorder,
we observe the predicted exotic strong-noise behavior to emerge after a large crossover time (see Fig.\ \ref{fig:Ps_t_weak_1d}) while the simulations
do not reach the asymptotic regime for even weaker disorder.
A positive confirmation of the exotic strong-noise critical point in weakly disordered systems
will therefore require a significantly higher numerical effort
\footnote{Distinguishing slowly varying functional forms such as logarithms and small powers
based on numerical data is notoriously difficult.}.

Let us compare our results to those of Jensen who applied Monte-Carlo simulations \cite{Jensen96}
and series expansions \cite{Jensen96,Jensen05} to directed bond percolation with temporal disorder in $1+1$ dimensions.
In contrast to the exotic behavior found in the present paper, Jensen reported a critical point with conventional power-law scaling,
but with nonuniversal critical exponents that change continuously with the disorder strength. What are the reasons for this
disagreement? The disorder considered in Refs.\ \cite{Jensen96,Jensen05} is not particularly strong. Based on the slow crossover
that we observed between the clean and disordered critical points, we believe that
Jensen's critical behavior may not be in the true asymptotic regime. This is supported by the fact that
some of the reported values for the correlation time exponent
$\nu_\parallel$ violate Kinzel's bound $\nu_\parallel > 2$. Alternatively, our strong-noise theory may hold only
for sufficiently strong disorder while weakly disordered systems display Jensen's nonuniversal power-law scaling.
(Note however, that we have observed deviations from power-law behavior even for weakly disordered
systems for which our simulations do not reach the asymptotic strong-noise regime.)

In addition to the critical behavior, we have also investigated the life time $\tau_N$ of finite-size samples
in the active phase (but close to criticality). The strong-noise RG predicts the existence
of temporal Griffiths phases which feature an anomalous power-law dependence,
$\tau_N \sim N^{1/\kappa}$, between life time and sample size (volume) $N$. It also predicts
that the Griffiths exponent $\kappa$ increases monotonically as the critical point is approached from
the active side, reaching the value $\kappa_c=d$ right at criticality. (In the mean-field limit
$d\to \infty$, $\kappa_c$ diverges, implying a logarithmic dependence of $\tau_N$ on $N$.)
Our Monte-Carlo simulations have demonstrated these temporal Griffiths phases in one and two dimensions.
In both cases, $\kappa$ varies with the infection rate as predicted, and the Monte-Carlo estimates
for $\kappa_c$
agree reasonably well with the RG prediction.

Thus, while our results confirm the notion of a temporal Griffiths phase introduced in
Ref.\ \cite{VBLM11}, the details are somewhat different. Reference \cite{VBLM11} did not find any anomalous
behavior of the life time in one dimension. This could be due to the very slow crossover from the
clean critical point to the true asymptotic behavior discussed at the end of Sec.\  \ref{sec:MC_1d}
which implies that simulations of weakly disordered systems may not reach the Griffiths regime within
achievable simulation times. Moreover, in two dimensions, Ref.\ \cite{VBLM11} reported
a logarithmic dependence of the life time at criticality on the system size, in contrast to our
power law with exponent $1/\kappa_c$. This difference could stem from the location of the critical point:
If the estimate used in Ref.\ \cite{VBLM11} was slightly on the inactive side of the transition,
a logarithmic dependence of the life time on the system size would naturally appear.

In recent years, contact processes on various types of complex networks has attracted significant attention
(see, e.g., Refs.\ \cite{CastellanoPastorSatorras06,NohPark09,JuhaszOdor09,FFPS11,FFCPS11,JOCM12,JuhaszKovacs13}).
It is interesting to ask whether temporal disorder is a relevant perturbation of the critical behavior of
these processes. The perturbative stability against temporal disorder should be governed by Kinzel's
criterion $\nu_\parallel > 2$. We believe our renormalization group theory can be generalized to this problem
by modifying the description of the spreading segments to account for the nontrivial
connectivity of the various networks. This remains a task for the future.

While clearcut experimental realizations of absorbing state phase transitions were missing for a long time,
they have recently been observed in turbulent states of certain liquid crystals \cite{TKCS07},
driven suspensions \cite{CCGP08,FFGP11}, the dynamics of superconducting vortices
\cite{OkumaTsugawaMotohashi11}, as well as in growing bacteria colonies \cite{KorolevNelson11,KXNF11}.
Investigating these transitions under the influence of external noise will permit experimental tests of our theory.
In particular, the effects of environmental fluctuations on the extinction of a biological population
or an entire biological species are attracting considerable attention in the context of global warming and other
large-scale environmental changes (see, e.g., Ref.\ \cite{OvaskainenMeerson10}).
In the laboratory, these questions could be
studied, e.g., by growing bacteria or yeast populations in fluctuating external conditions.

\section*{Acknowledgements}

This work was supported in part by the NSF under Grant Nos.\ DMR-1205803 and DMR-1506152,
by CNPq under Grant No.\ 307548/2015-5, and by FAPESP under Grant No.\ 2015/23849-7.

\appendix
\section{Time-reversal symmetry of the directed percolation universality class}
\label{sec:Appendix_A}

The DP universality class has a special symmetry under time reversal
\cite{GrassbergerdelaTorre79} that connects spreading and density decay experiments.
Because of this symmetry, the DP universality class is completely characterized by three
independent critical exponents rather than four (as is the case for a general absorbing
state transition).

\begin{figure}[b]
\includegraphics[width=8cm,clip]{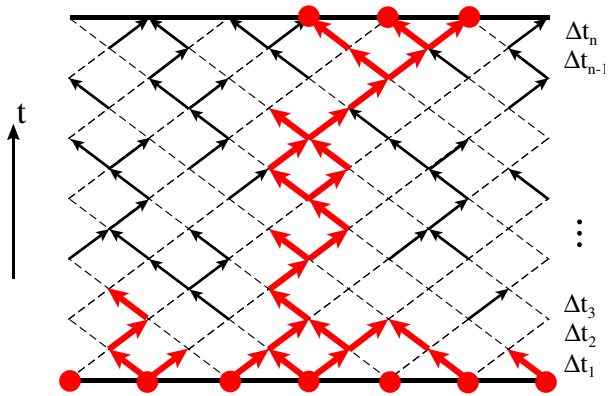}
\caption{(color online) Directed bond percolation in (1+1) dimensions. Arrows represent occupied bonds
while dashed lines indicate empty bonds.
A density decay experiment starts with all sites being active (red dots in the bottom row).
The red (thick) arrows show how the activity spreads with increasing time. }
\label{fig:DP}
\end{figure}

In this appendix, we demonstrate that the time-reversal symmetry still holds
(for disorder-averaged quantities) in the
presence of temporal disorder, generalizing arguments given in Ref.\ \cite{Hinrichsen00}.
Let us consider $(1+1)$-dimensional directed bond percolation.
Figure \ref{fig:DP} shows an example of a density decay experiment that begins
(bottom row) from a fully active lattice.
The density $\rho$ at the final time is given by the fraction of sites in the top row that are connected
via a directed path to at least one site in the bottom row.

If we now reverse all arrows in Fig.\ \ref{fig:DP}, we obtain a directed bond percolation process
running backwards in time (governed by the same bond occupation probabilities as the original
process).  If a lattice site in the top row was connected by a directed path
to the bottom row in the original process, it is also connected in the time reversed process.
A spreading experiment starting from any such site in the top row will therefore survive
to the bottom row. The survival probability $P_s$ is thus the fraction of sites in the top row with
directed connections to the bottom row. This is exactly the same as the density above, $P_s(t)=\rho(t)$.

These arguments establish that the density $\rho(t)$ for an individual realization of
the directed bond percolation process is identical to the survival probability $P_s(t)$ for the
corresponding realization with the same bond occupations but reversed order of the time steps $\Delta t_i$.
If the bond occupation probabilities do not depend on space and time (i.e., for the clean directed percolation problem),
these two realizations obviously occur with the same probability in the ensemble of all realizations of the
directed bond percolation process. After averaging over this ensemble, $\rho(t)$ and $P_s(t)$ are therefore
identical.
Importantly, even if the occupation probabilities themselves are disordered in space and/or time
this conclusion holds for the \emph{disorder averaged} $\rho(t)$ and $P_s(t)$
provided that the distributions of the occupation probabilities (the equivalent of $W_\lambda$ and $W_\mu$ in the main part of the article)
are time-independent
\footnote{Strictly, the distributions do not have to be \emph{time-independent}, they just have to be
invariant under time reversal.}.

The equivalence of $\rho$ and $P_s$ in higher-dimensional directed bond percolation can be
shown in the same way. For other microscopic realizations of the DP universality class,
$\rho$ and $P_s$ do not have to be identical. Universality guaranties, however, that they share the same
critical behavior.


\bibliography{../00Bibtex/rareregions}

\begin{thebibliography}{65}%
\makeatletter
\providecommand \@ifxundefined [1]{%
 \@ifx{#1\undefined}
}%
\providecommand \@ifnum [1]{%
 \ifnum #1\expandafter \@firstoftwo
 \else \expandafter \@secondoftwo
 \fi
}%
\providecommand \@ifx [1]{%
 \ifx #1\expandafter \@firstoftwo
 \else \expandafter \@secondoftwo
 \fi
}%
\providecommand \natexlab [1]{#1}%
\providecommand \enquote  [1]{``#1''}%
\providecommand \bibnamefont  [1]{#1}%
\providecommand \bibfnamefont [1]{#1}%
\providecommand \citenamefont [1]{#1}%
\providecommand \href@noop [0]{\@secondoftwo}%
\providecommand \href [0]{\begingroup \@sanitize@url \@href}%
\providecommand \@href[1]{\@@startlink{#1}\@@href}%
\providecommand \@@href[1]{\endgroup#1\@@endlink}%
\providecommand \@sanitize@url [0]{\catcode `\\12\catcode `\$12\catcode
  `\&12\catcode `\#12\catcode `\^12\catcode `\_12\catcode `\%12\relax}%
\providecommand \@@startlink[1]{}%
\providecommand \@@endlink[0]{}%
\providecommand \url  [0]{\begingroup\@sanitize@url \@url }%
\providecommand \@url [1]{\endgroup\@href {#1}{\urlprefix }}%
\providecommand \urlprefix  [0]{URL }%
\providecommand \Eprint [0]{\href }%
\providecommand \doibase [0]{http://dx.doi.org/}%
\providecommand \selectlanguage [0]{\@gobble}%
\providecommand \bibinfo  [0]{\@secondoftwo}%
\providecommand \bibfield  [0]{\@secondoftwo}%
\providecommand \translation [1]{[#1]}%
\providecommand \BibitemOpen [0]{}%
\providecommand \bibitemStop [0]{}%
\providecommand \bibitemNoStop [0]{.\EOS\space}%
\providecommand \EOS [0]{\spacefactor3000\relax}%
\providecommand \BibitemShut  [1]{\csname bibitem#1\endcsname}%
\let\auto@bib@innerbib\@empty
\bibitem [{\citenamefont {Janssen}(1981)}]{Janssen81}%
  \BibitemOpen
  \bibfield  {author} {\bibinfo {author} {\bibfnamefont {H.~K.}\ \bibnamefont
  {Janssen}},\ }\href@noop {} {\bibfield  {journal} {\bibinfo  {journal} {Z.
  Phys. B}\ }\textbf {\bibinfo {volume} {42}},\ \bibinfo {pages} {151}
  (\bibinfo {year} {1981})}\BibitemShut {NoStop}%
\bibitem [{\citenamefont {Grassberger}(1982)}]{Grassberger82}%
  \BibitemOpen
  \bibfield  {author} {\bibinfo {author} {\bibfnamefont {P.}~\bibnamefont
  {Grassberger}},\ }\href@noop {} {\bibfield  {journal} {\bibinfo  {journal}
  {Z. Phys. B}\ }\textbf {\bibinfo {volume} {47}},\ \bibinfo {pages} {365}
  (\bibinfo {year} {1982})}\BibitemShut {NoStop}%
\bibitem [{\citenamefont {Harris}(1974{\natexlab{a}})}]{HarrisTE74}%
  \BibitemOpen
  \bibfield  {author} {\bibinfo {author} {\bibfnamefont {T.~E.}\ \bibnamefont
  {Harris}},\ }\href {\doibase doi:10.1214/aop/1176996493} {\bibfield
  {journal} {\bibinfo  {journal} {Ann. Prob.}\ }\textbf {\bibinfo {volume}
  {2}},\ \bibinfo {pages} {969} (\bibinfo {year}
  {1974}{\natexlab{a}})}\BibitemShut {NoStop}%
\bibitem [{\citenamefont {Ziff}\ \emph {et~al.}(1986)\citenamefont {Ziff},
  \citenamefont {Gulari},\ and\ \citenamefont {Barshad}}]{ZiffGulariBarshad86}%
  \BibitemOpen
  \bibfield  {author} {\bibinfo {author} {\bibfnamefont {R.~M.}\ \bibnamefont
  {Ziff}}, \bibinfo {author} {\bibfnamefont {E.}~\bibnamefont {Gulari}}, \ and\
  \bibinfo {author} {\bibfnamefont {Y.}~\bibnamefont {Barshad}},\ }\href@noop
  {} {\bibfield  {journal} {\bibinfo  {journal} {Phys. Rev. Lett.}\ }\textbf
  {\bibinfo {volume} {56}},\ \bibinfo {pages} {2553} (\bibinfo {year}
  {1986})}\BibitemShut {NoStop}%
\bibitem [{\citenamefont {Tang}\ and\ \citenamefont
  {Leschhorn}(1992)}]{TangLeschhorn92}%
  \BibitemOpen
  \bibfield  {author} {\bibinfo {author} {\bibfnamefont {L.~H.}\ \bibnamefont
  {Tang}}\ and\ \bibinfo {author} {\bibfnamefont {H.}~\bibnamefont
  {Leschhorn}},\ }\href@noop {} {\bibfield  {journal} {\bibinfo  {journal}
  {Phys. Rev. A}\ }\textbf {\bibinfo {volume} {45}},\ \bibinfo {pages} {R8309}
  (\bibinfo {year} {1992})}\BibitemShut {NoStop}%
\bibitem [{\citenamefont {Barab\'asi}\ \emph {et~al.}(1996)\citenamefont
  {Barab\'asi}, \citenamefont {Grinstein},\ and\ \citenamefont
  {Mu\~noz}}]{BarabasiGrinsteinMunoz96}%
  \BibitemOpen
  \bibfield  {author} {\bibinfo {author} {\bibfnamefont {A.-L.}\ \bibnamefont
  {Barab\'asi}}, \bibinfo {author} {\bibfnamefont {G.}~\bibnamefont
  {Grinstein}}, \ and\ \bibinfo {author} {\bibfnamefont {M.~A.}\ \bibnamefont
  {Mu\~noz}},\ }\href {\doibase 10.1103/PhysRevLett.76.1481} {\bibfield
  {journal} {\bibinfo  {journal} {Phys. Rev. Lett.}\ }\textbf {\bibinfo
  {volume} {76}},\ \bibinfo {pages} {1481} (\bibinfo {year}
  {1996})}\BibitemShut {NoStop}%
\bibitem [{\citenamefont {Pomeau}(1986)}]{Pomeau86}%
  \BibitemOpen
  \bibfield  {author} {\bibinfo {author} {\bibfnamefont {Y.}~\bibnamefont
  {Pomeau}},\ }\href@noop {} {\bibfield  {journal} {\bibinfo  {journal}
  {Physica D}\ }\textbf {\bibinfo {volume} {23}},\ \bibinfo {pages} {3}
  (\bibinfo {year} {1986})}\BibitemShut {NoStop}%
\bibitem [{\citenamefont {Marro}\ and\ \citenamefont
  {Dickman}(1999)}]{MarroDickman99}%
  \BibitemOpen
  \bibfield  {author} {\bibinfo {author} {\bibfnamefont {J.}~\bibnamefont
  {Marro}}\ and\ \bibinfo {author} {\bibfnamefont {R.}~\bibnamefont
  {Dickman}},\ }\href@noop {} {\emph {\bibinfo {title} {Nonequilibrium Phase
  Transitions in Lattice Models}}}\ (\bibinfo  {publisher} {Cambridge
  University Press},\ \bibinfo {address} {Cambridge},\ \bibinfo {year}
  {1999})\BibitemShut {NoStop}%
\bibitem [{\citenamefont {Hinrichsen}(2000{\natexlab{a}})}]{Hinrichsen00}%
  \BibitemOpen
  \bibfield  {author} {\bibinfo {author} {\bibfnamefont {H.}~\bibnamefont
  {Hinrichsen}},\ }\href {\doibase 10.1080/00018730050198152} {\bibfield
  {journal} {\bibinfo  {journal} {Adv. Phys.}\ }\textbf {\bibinfo {volume}
  {49}},\ \bibinfo {pages} {815} (\bibinfo {year}
  {2000}{\natexlab{a}})}\BibitemShut {NoStop}%
\bibitem [{\citenamefont {Odor}(2004)}]{Odor04}%
  \BibitemOpen
  \bibfield  {author} {\bibinfo {author} {\bibfnamefont {G.}~\bibnamefont
  {Odor}},\ }\href {\doibase 10.1103/RevModPhys.76.663} {\bibfield  {journal}
  {\bibinfo  {journal} {Rev. Mod. Phys.}\ }\textbf {\bibinfo {volume} {76}},\
  \bibinfo {pages} {663} (\bibinfo {year} {2004})}\BibitemShut {NoStop}%
\bibitem [{\citenamefont {L{\"u}beck}(2004)}]{Luebeck04}%
  \BibitemOpen
  \bibfield  {author} {\bibinfo {author} {\bibfnamefont {S.}~\bibnamefont
  {L{\"u}beck}},\ }\href {\doibase 10.1142/S0217979204027748} {\bibfield
  {journal} {\bibinfo  {journal} {Int. J. Mod. Phys. B}\ }\textbf {\bibinfo
  {volume} {18}},\ \bibinfo {pages} {3977} (\bibinfo {year}
  {2004})}\BibitemShut {NoStop}%
\bibitem [{\citenamefont {T\"auber}\ \emph {et~al.}(2005)\citenamefont
  {T\"auber}, \citenamefont {Howard},\ and\ \citenamefont
  {Vollmayr-Lee}}]{TauberHowardVollmayrLee05}%
  \BibitemOpen
  \bibfield  {author} {\bibinfo {author} {\bibfnamefont {U.~C.}\ \bibnamefont
  {T\"auber}}, \bibinfo {author} {\bibfnamefont {M.}~\bibnamefont {Howard}}, \
  and\ \bibinfo {author} {\bibfnamefont {B.~P.}\ \bibnamefont {Vollmayr-Lee}},\
  }\href {http://stacks.iop.org/0305-4470/38/i=17/a=R01} {\bibfield  {journal}
  {\bibinfo  {journal} {J. Phys. A}\ }\textbf {\bibinfo {volume} {38}},\
  \bibinfo {pages} {R79} (\bibinfo {year} {2005})}\BibitemShut {NoStop}%
\bibitem [{\citenamefont {Henkel}\ \emph {et~al.}(2008)\citenamefont {Henkel},
  \citenamefont {Hinrichsen},\ and\ \citenamefont
  {L{\"u}beck}}]{HenkelHinrichsenLuebeck_book08}%
  \BibitemOpen
  \bibfield  {author} {\bibinfo {author} {\bibfnamefont {M.}~\bibnamefont
  {Henkel}}, \bibinfo {author} {\bibfnamefont {H.}~\bibnamefont {Hinrichsen}},
  \ and\ \bibinfo {author} {\bibfnamefont {S.}~\bibnamefont {L{\"u}beck}},\
  }\href@noop {} {\emph {\bibinfo {title} {Non-equilibrium phase transitions.
  Vol 1: Absorbing phase transitions}}}\ (\bibinfo  {publisher} {Springer},\
  \bibinfo {address} {Dordrecht},\ \bibinfo {year} {2008})\BibitemShut
  {NoStop}%
\bibitem [{\citenamefont {Hinrichsen}(2000{\natexlab{b}})}]{Hinrichsen00b}%
  \BibitemOpen
  \bibfield  {author} {\bibinfo {author} {\bibfnamefont {H.}~\bibnamefont
  {Hinrichsen}},\ }\href@noop {} {\bibfield  {journal} {\bibinfo  {journal}
  {Braz. J. Phys.}\ }\textbf {\bibinfo {volume} {30}},\ \bibinfo {pages} {69}
  (\bibinfo {year} {2000}{\natexlab{b}})}\BibitemShut {NoStop}%
\bibitem [{\citenamefont {Takeuchi}\ \emph {et~al.}(2007)\citenamefont
  {Takeuchi}, \citenamefont {Kuroda}, \citenamefont {Chat\'e},\ and\
  \citenamefont {Sano}}]{TKCS07}%
  \BibitemOpen
  \bibfield  {author} {\bibinfo {author} {\bibfnamefont {K.~A.}\ \bibnamefont
  {Takeuchi}}, \bibinfo {author} {\bibfnamefont {M.}~\bibnamefont {Kuroda}},
  \bibinfo {author} {\bibfnamefont {H.}~\bibnamefont {Chat\'e}}, \ and\
  \bibinfo {author} {\bibfnamefont {M.}~\bibnamefont {Sano}},\ }\href {\doibase
  10.1103/PhysRevLett.99.234503} {\bibfield  {journal} {\bibinfo  {journal}
  {Phys. Rev. Lett.}\ }\textbf {\bibinfo {volume} {99}},\ \bibinfo {pages}
  {234503} (\bibinfo {year} {2007})}\BibitemShut {NoStop}%
\bibitem [{\citenamefont {Corte}\ \emph {et~al.}(2008)\citenamefont {Corte},
  \citenamefont {Chaikin}, \citenamefont {Gollub},\ and\ \citenamefont
  {Pine}}]{CCGP08}%
  \BibitemOpen
  \bibfield  {author} {\bibinfo {author} {\bibfnamefont {L.}~\bibnamefont
  {Corte}}, \bibinfo {author} {\bibfnamefont {P.~M.}\ \bibnamefont {Chaikin}},
  \bibinfo {author} {\bibfnamefont {J.~P.}\ \bibnamefont {Gollub}}, \ and\
  \bibinfo {author} {\bibfnamefont {D.~J.}\ \bibnamefont {Pine}},\ }\href@noop
  {} {\bibfield  {journal} {\bibinfo  {journal} {Nature Physics}\ }\textbf
  {\bibinfo {volume} {4}},\ \bibinfo {pages} {420} (\bibinfo {year}
  {2008})}\BibitemShut {NoStop}%
\bibitem [{\citenamefont {Franceschini}\ \emph {et~al.}(2011)\citenamefont
  {Franceschini}, \citenamefont {Filippidi}, \citenamefont {Guazzelli},\ and\
  \citenamefont {Pine}}]{FFGP11}%
  \BibitemOpen
  \bibfield  {author} {\bibinfo {author} {\bibfnamefont {A.}~\bibnamefont
  {Franceschini}}, \bibinfo {author} {\bibfnamefont {E.}~\bibnamefont
  {Filippidi}}, \bibinfo {author} {\bibfnamefont {E.}~\bibnamefont
  {Guazzelli}}, \ and\ \bibinfo {author} {\bibfnamefont {D.~J.}\ \bibnamefont
  {Pine}},\ }\href {\doibase 10.1103/PhysRevLett.107.250603} {\bibfield
  {journal} {\bibinfo  {journal} {Phys. Rev. Lett.}\ }\textbf {\bibinfo
  {volume} {107}},\ \bibinfo {pages} {250603} (\bibinfo {year}
  {2011})}\BibitemShut {NoStop}%
\bibitem [{\citenamefont {Okuma}\ \emph {et~al.}(2011)\citenamefont {Okuma},
  \citenamefont {Tsugawa},\ and\ \citenamefont
  {Motohashi}}]{OkumaTsugawaMotohashi11}%
  \BibitemOpen
  \bibfield  {author} {\bibinfo {author} {\bibfnamefont {S.}~\bibnamefont
  {Okuma}}, \bibinfo {author} {\bibfnamefont {Y.}~\bibnamefont {Tsugawa}}, \
  and\ \bibinfo {author} {\bibfnamefont {A.}~\bibnamefont {Motohashi}},\ }\href
  {\doibase 10.1103/PhysRevB.83.012503} {\bibfield  {journal} {\bibinfo
  {journal} {Phys. Rev. B}\ }\textbf {\bibinfo {volume} {83}},\ \bibinfo
  {pages} {012503} (\bibinfo {year} {2011})}\BibitemShut {NoStop}%
\bibitem [{\citenamefont {Korolev}\ and\ \citenamefont
  {Nelson}(2011)}]{KorolevNelson11}%
  \BibitemOpen
  \bibfield  {author} {\bibinfo {author} {\bibfnamefont {K.~S.}\ \bibnamefont
  {Korolev}}\ and\ \bibinfo {author} {\bibfnamefont {D.~R.}\ \bibnamefont
  {Nelson}},\ }\href {\doibase 10.1103/PhysRevLett.107.088103} {\bibfield
  {journal} {\bibinfo  {journal} {Phys. Rev. Lett.}\ }\textbf {\bibinfo
  {volume} {107}},\ \bibinfo {pages} {088103} (\bibinfo {year}
  {2011})}\BibitemShut {NoStop}%
\bibitem [{\citenamefont {Korolev}\ \emph {et~al.}(2011)\citenamefont
  {Korolev}, \citenamefont {Xavier}, \citenamefont {Nelson},\ and\
  \citenamefont {Foster}}]{KXNF11}%
  \BibitemOpen
  \bibfield  {author} {\bibinfo {author} {\bibfnamefont {K.~S.}\ \bibnamefont
  {Korolev}}, \bibinfo {author} {\bibfnamefont {J.~B.}\ \bibnamefont {Xavier}},
  \bibinfo {author} {\bibfnamefont {D.~R.}\ \bibnamefont {Nelson}}, \ and\
  \bibinfo {author} {\bibfnamefont {K.~R.}\ \bibnamefont {Foster}},\
  }\href@noop {} {\bibfield  {journal} {\bibinfo  {journal} {The American
  Naturalist}\ }\textbf {\bibinfo {volume} {178}},\ \bibinfo {pages} {538}
  (\bibinfo {year} {2011})}\BibitemShut {NoStop}%
\bibitem [{\citenamefont {Harris}(1974{\natexlab{b}})}]{Harris74}%
  \BibitemOpen
  \bibfield  {author} {\bibinfo {author} {\bibfnamefont {A.~B.}\ \bibnamefont
  {Harris}},\ }\href {\doibase 10.1088/0022-3719/7/9/009} {\bibfield  {journal}
  {\bibinfo  {journal} {J. Phys. C}\ }\textbf {\bibinfo {volume} {7}},\
  \bibinfo {pages} {1671} (\bibinfo {year} {1974}{\natexlab{b}})}\BibitemShut
  {NoStop}%
\bibitem [{\citenamefont {Kinzel}(1985)}]{Kinzel85}%
  \BibitemOpen
  \bibfield  {author} {\bibinfo {author} {\bibfnamefont {W.}~\bibnamefont
  {Kinzel}},\ }\href@noop {} {\bibfield  {journal} {\bibinfo  {journal} {Z.
  Phys. B}\ }\textbf {\bibinfo {volume} {58}},\ \bibinfo {pages} {229}
  (\bibinfo {year} {1985})}\BibitemShut {NoStop}%
\bibitem [{\citenamefont {Vojta}\ and\ \citenamefont
  {Dickman}(2016)}]{VojtaDickman16}%
  \BibitemOpen
  \bibfield  {author} {\bibinfo {author} {\bibfnamefont {T.}~\bibnamefont
  {Vojta}}\ and\ \bibinfo {author} {\bibfnamefont {R.}~\bibnamefont
  {Dickman}},\ }\href {\doibase 10.1103/PhysRevE.93.032143} {\bibfield
  {journal} {\bibinfo  {journal} {Phys. Rev. E}\ }\textbf {\bibinfo {volume}
  {93}},\ \bibinfo {pages} {032143} (\bibinfo {year} {2016})}\BibitemShut
  {NoStop}%
\bibitem [{\citenamefont {Hooyberghs}\ \emph {et~al.}(2003)\citenamefont
  {Hooyberghs}, \citenamefont {Igl\'oi},\ and\ \citenamefont
  {Vanderzande}}]{HooyberghsIgloiVanderzande03}%
  \BibitemOpen
  \bibfield  {author} {\bibinfo {author} {\bibfnamefont {J.}~\bibnamefont
  {Hooyberghs}}, \bibinfo {author} {\bibfnamefont {F.}~\bibnamefont {Igl\'oi}},
  \ and\ \bibinfo {author} {\bibfnamefont {C.}~\bibnamefont {Vanderzande}},\
  }\href {\doibase 10.1103/PhysRevLett.90.100601} {\bibfield  {journal}
  {\bibinfo  {journal} {Phys. Rev. Lett.}\ }\textbf {\bibinfo {volume} {90}},\
  \bibinfo {pages} {100601} (\bibinfo {year} {2003})}\BibitemShut {NoStop}%
\bibitem [{\citenamefont {Hooyberghs}\ \emph {et~al.}(2004)\citenamefont
  {Hooyberghs}, \citenamefont {Igl\'oi},\ and\ \citenamefont
  {Vanderzande}}]{HooyberghsIgloiVanderzande04}%
  \BibitemOpen
  \bibfield  {author} {\bibinfo {author} {\bibfnamefont {J.}~\bibnamefont
  {Hooyberghs}}, \bibinfo {author} {\bibfnamefont {F.}~\bibnamefont {Igl\'oi}},
  \ and\ \bibinfo {author} {\bibfnamefont {C.}~\bibnamefont {Vanderzande}},\
  }\href {\doibase 10.1103/PhysRevE.69.066140} {\bibfield  {journal} {\bibinfo
  {journal} {Phys. Rev. E}\ }\textbf {\bibinfo {volume} {69}},\ \bibinfo
  {pages} {066140} (\bibinfo {year} {2004})}\BibitemShut {NoStop}%
\bibitem [{\citenamefont {Ma}\ \emph {et~al.}(1979)\citenamefont {Ma},
  \citenamefont {Dasgupta},\ and\ \citenamefont {Hu}}]{MaDasguptaHu79}%
  \BibitemOpen
  \bibfield  {author} {\bibinfo {author} {\bibfnamefont {S.~K.}\ \bibnamefont
  {Ma}}, \bibinfo {author} {\bibfnamefont {C.}~\bibnamefont {Dasgupta}}, \ and\
  \bibinfo {author} {\bibfnamefont {C.~K.}\ \bibnamefont {Hu}},\ }\href@noop {}
  {\bibfield  {journal} {\bibinfo  {journal} {Phys. Rev. Lett.}\ }\textbf
  {\bibinfo {volume} {43}},\ \bibinfo {pages} {1434} (\bibinfo {year}
  {1979})}\BibitemShut {NoStop}%
\bibitem [{\citenamefont {Igloi}\ and\ \citenamefont
  {Monthus}(2005)}]{IgloiMonthus05}%
  \BibitemOpen
  \bibfield  {author} {\bibinfo {author} {\bibfnamefont {F.}~\bibnamefont
  {Igloi}}\ and\ \bibinfo {author} {\bibfnamefont {C.}~\bibnamefont
  {Monthus}},\ }\href@noop {} {\bibfield  {journal} {\bibinfo  {journal} {Phys.
  Rep.}\ }\textbf {\bibinfo {volume} {412}},\ \bibinfo {pages} {277} (\bibinfo
  {year} {2005})}\BibitemShut {NoStop}%
\bibitem [{Note1()}]{Note1}%
  \BibitemOpen
  \bibinfo {note} {A self-consistent extension of the method to the
  weak-disorder regime is presented in Ref.\ \cite {Hoyos08}.}\BibitemShut
  {Stop}%
\bibitem [{\citenamefont {Griffiths}(1969)}]{Griffiths69}%
  \BibitemOpen
  \bibfield  {author} {\bibinfo {author} {\bibfnamefont {R.~B.}\ \bibnamefont
  {Griffiths}},\ }\href {\doibase 10.1103/PhysRevLett.23.17} {\bibfield
  {journal} {\bibinfo  {journal} {Phys. Rev. Lett.}\ }\textbf {\bibinfo
  {volume} {23}},\ \bibinfo {pages} {17} (\bibinfo {year} {1969})}\BibitemShut
  {NoStop}%
\bibitem [{\citenamefont {Noest}(1986)}]{Noest86}%
  \BibitemOpen
  \bibfield  {author} {\bibinfo {author} {\bibfnamefont {A.~J.}\ \bibnamefont
  {Noest}},\ }\href {\doibase 10.1103/PhysRevLett.57.90} {\bibfield  {journal}
  {\bibinfo  {journal} {Phys. Rev. Lett.}\ }\textbf {\bibinfo {volume} {57}},\
  \bibinfo {pages} {90} (\bibinfo {year} {1986})}\BibitemShut {NoStop}%
\bibitem [{\citenamefont {Noest}(1988)}]{Noest88}%
  \BibitemOpen
  \bibfield  {author} {\bibinfo {author} {\bibfnamefont {A.~J.}\ \bibnamefont
  {Noest}},\ }\href {\doibase 10.1103/PhysRevB.38.2715} {\bibfield  {journal}
  {\bibinfo  {journal} {Phys. Rev. B}\ }\textbf {\bibinfo {volume} {38}},\
  \bibinfo {pages} {2715} (\bibinfo {year} {1988})}\BibitemShut {NoStop}%
\bibitem [{\citenamefont {Vojta}\ and\ \citenamefont
  {Dickison}(2005)}]{VojtaDickison05}%
  \BibitemOpen
  \bibfield  {author} {\bibinfo {author} {\bibfnamefont {T.}~\bibnamefont
  {Vojta}}\ and\ \bibinfo {author} {\bibfnamefont {M.}~\bibnamefont
  {Dickison}},\ }\href {\doibase 10.1103/PhysRevE.72.036126} {\bibfield
  {journal} {\bibinfo  {journal} {Phys. Rev. E}\ }\textbf {\bibinfo {volume}
  {72}},\ \bibinfo {pages} {036126} (\bibinfo {year} {2005})}\BibitemShut
  {NoStop}%
\bibitem [{\citenamefont {de~Oliveira}\ and\ \citenamefont
  {Ferreira}(2008)}]{OliveiraFerreira08}%
  \BibitemOpen
  \bibfield  {author} {\bibinfo {author} {\bibfnamefont {M.~M.}\ \bibnamefont
  {de~Oliveira}}\ and\ \bibinfo {author} {\bibfnamefont {S.~C.}\ \bibnamefont
  {Ferreira}},\ }\href@noop {} {\bibfield  {journal} {\bibinfo  {journal} {J.
  Stat. Mech.}\ }\textbf {\bibinfo {volume} {2008}},\ \bibinfo {pages} {P11001}
  (\bibinfo {year} {2008})}\BibitemShut {NoStop}%
\bibitem [{\citenamefont {Vojta}\ \emph {et~al.}(2009)\citenamefont {Vojta},
  \citenamefont {Farquhar},\ and\ \citenamefont {Mast}}]{VojtaFarquharMast09}%
  \BibitemOpen
  \bibfield  {author} {\bibinfo {author} {\bibfnamefont {T.}~\bibnamefont
  {Vojta}}, \bibinfo {author} {\bibfnamefont {A.}~\bibnamefont {Farquhar}}, \
  and\ \bibinfo {author} {\bibfnamefont {J.}~\bibnamefont {Mast}},\ }\href
  {\doibase 10.1103/PhysRevE.79.011111} {\bibfield  {journal} {\bibinfo
  {journal} {Phys. Rev. E}\ }\textbf {\bibinfo {volume} {79}},\ \bibinfo
  {pages} {011111} (\bibinfo {year} {2009})}\BibitemShut {NoStop}%
\bibitem [{\citenamefont {Vojta}(2012)}]{Vojta12}%
  \BibitemOpen
  \bibfield  {author} {\bibinfo {author} {\bibfnamefont {T.}~\bibnamefont
  {Vojta}},\ }\href {\doibase 10.1103/PhysRevE.86.051137} {\bibfield  {journal}
  {\bibinfo  {journal} {Phys. Rev. E}\ }\textbf {\bibinfo {volume} {86}},\
  \bibinfo {pages} {051137} (\bibinfo {year} {2012})}\BibitemShut {NoStop}%
\bibitem [{\citenamefont {Vojta}\ and\ \citenamefont {Lee}(2006)}]{VojtaLee06}%
  \BibitemOpen
  \bibfield  {author} {\bibinfo {author} {\bibfnamefont {T.}~\bibnamefont
  {Vojta}}\ and\ \bibinfo {author} {\bibfnamefont {M.~Y.}\ \bibnamefont
  {Lee}},\ }\href {\doibase 10.1103/PhysRevLett.96.035701} {\bibfield
  {journal} {\bibinfo  {journal} {Phys. Rev. Lett.}\ }\textbf {\bibinfo
  {volume} {96}},\ \bibinfo {pages} {035701} (\bibinfo {year}
  {2006})}\BibitemShut {NoStop}%
\bibitem [{\citenamefont {Lee}\ and\ \citenamefont {Vojta}(2009)}]{LeeVojta09}%
  \BibitemOpen
  \bibfield  {author} {\bibinfo {author} {\bibfnamefont {M.~Y.}\ \bibnamefont
  {Lee}}\ and\ \bibinfo {author} {\bibfnamefont {T.}~\bibnamefont {Vojta}},\
  }\href {\doibase 10.1103/PhysRevE.79.041112} {\bibfield  {journal} {\bibinfo
  {journal} {Phys. Rev. E}\ }\textbf {\bibinfo {volume} {79}},\ \bibinfo
  {pages} {041112} (\bibinfo {year} {2009})}\BibitemShut {NoStop}%
\bibitem [{\citenamefont {Barghathi}\ \emph {et~al.}(2014)\citenamefont
  {Barghathi}, \citenamefont {Nozadze},\ and\ \citenamefont
  {Vojta}}]{BarghathiNozadzeVojta14}%
  \BibitemOpen
  \bibfield  {author} {\bibinfo {author} {\bibfnamefont {H.}~\bibnamefont
  {Barghathi}}, \bibinfo {author} {\bibfnamefont {D.}~\bibnamefont {Nozadze}},
  \ and\ \bibinfo {author} {\bibfnamefont {T.}~\bibnamefont {Vojta}},\ }\href
  {\doibase 10.1103/PhysRevE.89.012112} {\bibfield  {journal} {\bibinfo
  {journal} {Phys. Rev. E}\ }\textbf {\bibinfo {volume} {89}},\ \bibinfo
  {pages} {012112} (\bibinfo {year} {2014})}\BibitemShut {NoStop}%
\bibitem [{\citenamefont {Vojta}(2006)}]{Vojta06}%
  \BibitemOpen
  \bibfield  {author} {\bibinfo {author} {\bibfnamefont {T.}~\bibnamefont
  {Vojta}},\ }\href {\doibase 10.1088/0305-4470/39/22/R01} {\bibfield
  {journal} {\bibinfo  {journal} {J. Phys. A}\ }\textbf {\bibinfo {volume}
  {39}},\ \bibinfo {pages} {R143} (\bibinfo {year} {2006})}\BibitemShut
  {NoStop}%
\bibitem [{\citenamefont {Jensen}(1996)}]{Jensen96}%
  \BibitemOpen
  \bibfield  {author} {\bibinfo {author} {\bibfnamefont {I.}~\bibnamefont
  {Jensen}},\ }\href {\doibase 10.1103/PhysRevLett.77.4988} {\bibfield
  {journal} {\bibinfo  {journal} {Phys. Rev. Lett.}\ }\textbf {\bibinfo
  {volume} {77}},\ \bibinfo {pages} {4988} (\bibinfo {year}
  {1996})}\BibitemShut {NoStop}%
\bibitem [{\citenamefont {Jensen}(2005)}]{Jensen05}%
  \BibitemOpen
  \bibfield  {author} {\bibinfo {author} {\bibfnamefont {I.}~\bibnamefont
  {Jensen}},\ }\href {http://stacks.iop.org/0305-4470/38/i=7/a=003} {\bibfield
  {journal} {\bibinfo  {journal} {J. Phys. A}\ }\textbf {\bibinfo {volume}
  {38}},\ \bibinfo {pages} {1441} (\bibinfo {year} {2005})}\BibitemShut
  {NoStop}%
\bibitem [{\citenamefont {Vazquez}\ \emph {et~al.}(2011)\citenamefont
  {Vazquez}, \citenamefont {Bonachela}, \citenamefont {L\'opez},\ and\
  \citenamefont {Mu\~noz}}]{VBLM11}%
  \BibitemOpen
  \bibfield  {author} {\bibinfo {author} {\bibfnamefont {F.}~\bibnamefont
  {Vazquez}}, \bibinfo {author} {\bibfnamefont {J.~A.}\ \bibnamefont
  {Bonachela}}, \bibinfo {author} {\bibfnamefont {C.}~\bibnamefont {L\'opez}},
  \ and\ \bibinfo {author} {\bibfnamefont {M.~A.}\ \bibnamefont {Mu\~noz}},\
  }\href {\doibase 10.1103/PhysRevLett.106.235702} {\bibfield  {journal}
  {\bibinfo  {journal} {Phys. Rev. Lett.}\ }\textbf {\bibinfo {volume} {106}},\
  \bibinfo {pages} {235702} (\bibinfo {year} {2011})}\BibitemShut {NoStop}%
\bibitem [{\citenamefont {Vojta}\ and\ \citenamefont
  {Hoyos}(2015)}]{VojtaHoyos15}%
  \BibitemOpen
  \bibfield  {author} {\bibinfo {author} {\bibfnamefont {T.}~\bibnamefont
  {Vojta}}\ and\ \bibinfo {author} {\bibfnamefont {J.~A.}\ \bibnamefont
  {Hoyos}},\ }\href {http://stacks.iop.org/0295-5075/112/i=3/a=30002}
  {\bibfield  {journal} {\bibinfo  {journal} {EPL (Europhysics Letters)}\
  }\textbf {\bibinfo {volume} {112}},\ \bibinfo {pages} {30002} (\bibinfo
  {year} {2015})}\BibitemShut {NoStop}%
\bibitem [{\citenamefont {Fisher}(1992)}]{Fisher92}%
  \BibitemOpen
  \bibfield  {author} {\bibinfo {author} {\bibfnamefont {D.~S.}\ \bibnamefont
  {Fisher}},\ }\href@noop {} {\bibfield  {journal} {\bibinfo  {journal} {Phys.
  Rev. Lett.}\ }\textbf {\bibinfo {volume} {69}},\ \bibinfo {pages} {534}
  (\bibinfo {year} {1992})}\BibitemShut {NoStop}%
\bibitem [{\citenamefont {Fisher}(1995)}]{Fisher95}%
  \BibitemOpen
  \bibfield  {author} {\bibinfo {author} {\bibfnamefont {D.~S.}\ \bibnamefont
  {Fisher}},\ }\href {\doibase 10.1103/PhysRevB.51.6411} {\bibfield  {journal}
  {\bibinfo  {journal} {Phys. Rev. B}\ }\textbf {\bibinfo {volume} {51}},\
  \bibinfo {pages} {6411} (\bibinfo {year} {1995})}\BibitemShut {NoStop}%
\bibitem [{\citenamefont {Vojta}\ \emph {et~al.}(2011)\citenamefont {Vojta},
  \citenamefont {Hoyos}, \citenamefont {Mohan},\ and\ \citenamefont
  {Narayanan}}]{VHMN11}%
  \BibitemOpen
  \bibfield  {author} {\bibinfo {author} {\bibfnamefont {T.}~\bibnamefont
  {Vojta}}, \bibinfo {author} {\bibfnamefont {J.~A.}\ \bibnamefont {Hoyos}},
  \bibinfo {author} {\bibfnamefont {P.}~\bibnamefont {Mohan}}, \ and\ \bibinfo
  {author} {\bibfnamefont {R.}~\bibnamefont {Narayanan}},\ }\href
  {http://stacks.iop.org/0953-8984/23/i=9/a=094206} {\bibfield  {journal}
  {\bibinfo  {journal} {J. Phys. Condens. Mat.}\ }\textbf {\bibinfo {volume}
  {23}},\ \bibinfo {pages} {094206} (\bibinfo {year} {2011})}\BibitemShut
  {NoStop}%
\bibitem [{\citenamefont {Juh\'asz}\ \emph {et~al.}(2014)\citenamefont
  {Juh\'asz}, \citenamefont {Kov\'acs},\ and\ \citenamefont
  {Igloi}}]{JuhaszKovacsIgloi14}%
  \BibitemOpen
  \bibfield  {author} {\bibinfo {author} {\bibfnamefont {R.}~\bibnamefont
  {Juh\'asz}}, \bibinfo {author} {\bibfnamefont {I.~A.}\ \bibnamefont
  {Kov\'acs}}, \ and\ \bibinfo {author} {\bibfnamefont {F.}~\bibnamefont
  {Igloi}},\ }\href {http://stacks.iop.org/0295-5075/107/i=4/a=47008}
  {\bibfield  {journal} {\bibinfo  {journal} {EPL (Europhysics Letters)}\
  }\textbf {\bibinfo {volume} {107}},\ \bibinfo {pages} {47008} (\bibinfo
  {year} {2014})}\BibitemShut {NoStop}%
\bibitem [{Note2()}]{Note2}%
  \BibitemOpen
  \bibinfo {note} {The exponential ansatz can be motivated by the fact that its
  functional form is invariant under the convolution operation in Eqs.\ (\ref
  {eq:flow-R}) and (\ref {eq:flow-P_finite_d}). For the flow equations arising
  in the mean-field case, Fisher \cite {Fisher92,*Fisher95} showed
  analytically, that the fixed point solutions must have this form unless the
  bare disorder distributions are highly singular. In the case of Eqs.\ (\ref
  {eq:flow-R}) and (\ref {eq:flow-P_finite_d}), the same has been shown by
  numerically iterating the RG recursion relations \cite
  {JuhaszKovacsIgloi14}.}\BibitemShut {Stop}%
\bibitem [{\citenamefont {Kosterlitz}\ and\ \citenamefont
  {Thouless}(1973)}]{KosterlitzThouless73}%
  \BibitemOpen
  \bibfield  {author} {\bibinfo {author} {\bibfnamefont {J.~M.}\ \bibnamefont
  {Kosterlitz}}\ and\ \bibinfo {author} {\bibfnamefont {D.~J.}\ \bibnamefont
  {Thouless}},\ }\href@noop {} {\bibfield  {journal} {\bibinfo  {journal} {J.
  Phys. C}\ }\textbf {\bibinfo {volume} {6}},\ \bibinfo {pages} {1181}
  (\bibinfo {year} {1973})}\BibitemShut {NoStop}%
\bibitem [{\citenamefont {Grassberger}\ and\ \citenamefont {de~la
  Torre}(1979)}]{GrassbergerdelaTorre79}%
  \BibitemOpen
  \bibfield  {author} {\bibinfo {author} {\bibfnamefont {P.}~\bibnamefont
  {Grassberger}}\ and\ \bibinfo {author} {\bibfnamefont {A.}~\bibnamefont
  {de~la Torre}},\ }\href {\doibase 10.1016/0003-4916(79)90207-0} {\bibfield
  {journal} {\bibinfo  {journal} {Ann. Phys. (NY)}\ }\textbf {\bibinfo {volume}
  {122}},\ \bibinfo {pages} {373} (\bibinfo {year} {1979})}\BibitemShut
  {NoStop}%
\bibitem [{\citenamefont {Dickman}(1999)}]{Dickman99}%
  \BibitemOpen
  \bibfield  {author} {\bibinfo {author} {\bibfnamefont {R.}~\bibnamefont
  {Dickman}},\ }\href@noop {} {\bibfield  {journal} {\bibinfo  {journal} {Phys.
  Rev. E}\ }\textbf {\bibinfo {volume} {60}},\ \bibinfo {pages} {R2441}
  (\bibinfo {year} {1999})}\BibitemShut {NoStop}%
\bibitem [{Note3()}]{Note3}%
  \BibitemOpen
  \bibinfo {note} {For weak disorder, these deviations are subtle and only
  visible in high-precision data.}\BibitemShut {Stop}%
\bibitem [{\citenamefont {Jensen}(1999)}]{Jensen99}%
  \BibitemOpen
  \bibfield  {author} {\bibinfo {author} {\bibfnamefont {I.}~\bibnamefont
  {Jensen}},\ }\href@noop {} {\bibfield  {journal} {\bibinfo  {journal} {J.
  Phys. A}\ }\textbf {\bibinfo {volume} {32}},\ \bibinfo {pages} {5233}
  (\bibinfo {year} {1999})}\BibitemShut {NoStop}%
\bibitem [{Note4()}]{Note4}%
  \BibitemOpen
  \bibinfo {note} {If the only relevant effect of spatial fluctuations was to
  change the time dependence of the density during spreading segments from
  ballistic to a weaker power law, $\rho (t) \approx \rho _0 (1+b t)^{d'}$ with
  $d'<d$, our theory would change very little. The recursion for $1/ \protect
  \mathaccentV {tilde}07Ea^\protect \mathrm {dn}$ would take the additive form
  (\ref {eq:tilde_a_dn_fd}) with $d$ replaced by $d'$. The resulting critical
  behavior \cite {VHMN11} would again be of Kosterlitz-Thouless form with the
  critical exponents given by (\ref {eq:finite-d-exponents}) except the
  dynamical exponent which would take the value $z=d/d'$.}\BibitemShut {Stop}%
\bibitem [{Note5()}]{Note5}%
  \BibitemOpen
  \bibinfo {note} {Distinguishing slowly varying functional forms such as
  logarithms and small powers based on numerical data is notoriously
  difficult.}\BibitemShut {Stop}%
\bibitem [{\citenamefont {Castellano}\ and\ \citenamefont
  {Pastor-Satorras}(2006)}]{CastellanoPastorSatorras06}%
  \BibitemOpen
  \bibfield  {author} {\bibinfo {author} {\bibfnamefont {C.}~\bibnamefont
  {Castellano}}\ and\ \bibinfo {author} {\bibfnamefont {R.}~\bibnamefont
  {Pastor-Satorras}},\ }\href {\doibase 10.1103/PhysRevLett.96.038701}
  {\bibfield  {journal} {\bibinfo  {journal} {Phys. Rev. Lett.}\ }\textbf
  {\bibinfo {volume} {96}},\ \bibinfo {pages} {038701} (\bibinfo {year}
  {2006})}\BibitemShut {NoStop}%
\bibitem [{\citenamefont {Noh}\ and\ \citenamefont {Park}(2009)}]{NohPark09}%
  \BibitemOpen
  \bibfield  {author} {\bibinfo {author} {\bibfnamefont {J.~D.}\ \bibnamefont
  {Noh}}\ and\ \bibinfo {author} {\bibfnamefont {H.}~\bibnamefont {Park}},\
  }\href {\doibase 10.1103/PhysRevE.79.056115} {\bibfield  {journal} {\bibinfo
  {journal} {Phys. Rev. E}\ }\textbf {\bibinfo {volume} {79}},\ \bibinfo
  {pages} {056115} (\bibinfo {year} {2009})}\BibitemShut {NoStop}%
\bibitem [{\citenamefont {Juh\'asz}\ and\ \citenamefont
  {\'Odor}(2009)}]{JuhaszOdor09}%
  \BibitemOpen
  \bibfield  {author} {\bibinfo {author} {\bibfnamefont {R.}~\bibnamefont
  {Juh\'asz}}\ and\ \bibinfo {author} {\bibfnamefont {G.}~\bibnamefont
  {\'Odor}},\ }\href {\doibase 10.1103/PhysRevE.80.041123} {\bibfield
  {journal} {\bibinfo  {journal} {Phys. Rev. E}\ }\textbf {\bibinfo {volume}
  {80}},\ \bibinfo {pages} {041123} (\bibinfo {year} {2009})}\BibitemShut
  {NoStop}%
\bibitem [{\citenamefont {Ferreira}\ \emph
  {et~al.}(2011{\natexlab{a}})\citenamefont {Ferreira}, \citenamefont
  {Ferreira},\ and\ \citenamefont {Pastor-Satorras}}]{FFPS11}%
  \BibitemOpen
  \bibfield  {author} {\bibinfo {author} {\bibfnamefont {S.~C.}\ \bibnamefont
  {Ferreira}}, \bibinfo {author} {\bibfnamefont {R.~S.}\ \bibnamefont
  {Ferreira}}, \ and\ \bibinfo {author} {\bibfnamefont {R.}~\bibnamefont
  {Pastor-Satorras}},\ }\href {\doibase 10.1103/PhysRevE.83.066113} {\bibfield
  {journal} {\bibinfo  {journal} {Phys. Rev. E}\ }\textbf {\bibinfo {volume}
  {83}},\ \bibinfo {pages} {066113} (\bibinfo {year}
  {2011}{\natexlab{a}})}\BibitemShut {NoStop}%
\bibitem [{\citenamefont {Ferreira}\ \emph
  {et~al.}(2011{\natexlab{b}})\citenamefont {Ferreira}, \citenamefont
  {Ferreira}, \citenamefont {Castellano},\ and\ \citenamefont
  {Pastor-Satorras}}]{FFCPS11}%
  \BibitemOpen
  \bibfield  {author} {\bibinfo {author} {\bibfnamefont {S.~C.}\ \bibnamefont
  {Ferreira}}, \bibinfo {author} {\bibfnamefont {R.~S.}\ \bibnamefont
  {Ferreira}}, \bibinfo {author} {\bibfnamefont {C.}~\bibnamefont
  {Castellano}}, \ and\ \bibinfo {author} {\bibfnamefont {R.}~\bibnamefont
  {Pastor-Satorras}},\ }\href {\doibase 10.1103/PhysRevE.84.066102} {\bibfield
  {journal} {\bibinfo  {journal} {Phys. Rev. E}\ }\textbf {\bibinfo {volume}
  {84}},\ \bibinfo {pages} {066102} (\bibinfo {year}
  {2011}{\natexlab{b}})}\BibitemShut {NoStop}%
\bibitem [{\citenamefont {Juh\'asz}\ \emph {et~al.}(2012)\citenamefont
  {Juh\'asz}, \citenamefont {\'Odor}, \citenamefont {Castellano},\ and\
  \citenamefont {Mu\~noz}}]{JOCM12}%
  \BibitemOpen
  \bibfield  {author} {\bibinfo {author} {\bibfnamefont {R.}~\bibnamefont
  {Juh\'asz}}, \bibinfo {author} {\bibfnamefont {G.}~\bibnamefont {\'Odor}},
  \bibinfo {author} {\bibfnamefont {C.}~\bibnamefont {Castellano}}, \ and\
  \bibinfo {author} {\bibfnamefont {M.~A.}\ \bibnamefont {Mu\~noz}},\ }\href
  {\doibase 10.1103/PhysRevE.85.066125} {\bibfield  {journal} {\bibinfo
  {journal} {Phys. Rev. E}\ }\textbf {\bibinfo {volume} {85}},\ \bibinfo
  {pages} {066125} (\bibinfo {year} {2012})}\BibitemShut {NoStop}%
\bibitem [{\citenamefont {Juh\'asz}\ and\ \citenamefont
  {Kov\'acs}(2013)}]{JuhaszKovacs13}%
  \BibitemOpen
  \bibfield  {author} {\bibinfo {author} {\bibfnamefont {R.}~\bibnamefont
  {Juh\'asz}}\ and\ \bibinfo {author} {\bibfnamefont {I.~A.}\ \bibnamefont
  {Kov\'acs}},\ }\href {http://stacks.iop.org/1742-5468/2013/i=06/a=P06003}
  {\bibfield  {journal} {\bibinfo  {journal} {J. Stat. Mech.}\ }\textbf
  {\bibinfo {volume} {2013}},\ \bibinfo {pages} {P06003} (\bibinfo {year}
  {2013})}\BibitemShut {NoStop}%
\bibitem [{\citenamefont {Ovaskainen}\ and\ \citenamefont
  {Meerson}(2010)}]{OvaskainenMeerson10}%
  \BibitemOpen
  \bibfield  {author} {\bibinfo {author} {\bibfnamefont {O.}~\bibnamefont
  {Ovaskainen}}\ and\ \bibinfo {author} {\bibfnamefont {B.}~\bibnamefont
  {Meerson}},\ }\href {\doibase http://dx.doi.org/10.1016/j.tree.2010.07.009}
  {\bibfield  {journal} {\bibinfo  {journal} {Trends in Ecology \& Evolution}\
  }\textbf {\bibinfo {volume} {25}},\ \bibinfo {pages} {643 } (\bibinfo {year}
  {2010})}\BibitemShut {NoStop}%
\bibitem [{Note6()}]{Note6}%
  \BibitemOpen
  \bibinfo {note} {Strictly, the distributions do not have to be \protect \emph
  {time-independent}, they just have to be invariant under time
  reversal.}\BibitemShut {Stop}%
\bibitem [{\citenamefont {Hoyos}(2008)}]{Hoyos08}%
  \BibitemOpen
  \bibfield  {author} {\bibinfo {author} {\bibfnamefont {J.~A.}\ \bibnamefont
  {Hoyos}},\ }\href {\doibase 10.1103/PhysRevE.78.032101} {\bibfield  {journal}
  {\bibinfo  {journal} {Phys. Rev. E}\ }\textbf {\bibinfo {volume} {78}},\
  \bibinfo {pages} {032101} (\bibinfo {year} {2008})}\BibitemShut {NoStop}%
\end{thebibliography}%
\end{document}